\documentclass[10pt]{amsart}

\usepackage{lipsum}
\usepackage{amsfonts,bm,amssymb,amsmath}
\usepackage{graphicx}
\usepackage{epstopdf}
\usepackage[caption=false]{subfig}
\usepackage{pgfplots}
\usepackage{algorithmic}
\usepackage{amsopn}
\usepackage{amsrefs}
\usepackage{float}
\usepackage{caption}

\usepackage{hyperref}
\usepackage{cleveref}

\usepackage{color}
\usepackage{diagbox}
\usepackage{geometry}

\newtheorem{remark}{Remark}[section]

\hypersetup{
	colorlinks,
	linkcolor={red!50!black},
	citecolor={blue!50!black},
	urlcolor={blue!80!black}
}

\numberwithin{equation}{section}

\definecolor{newcolor1}{rgb}{.8,.349,.1}
\colorlet{bblue}{blue!50!black}
\crefformat{equation}{(#2#1#3)}
\crefmultiformat{equation}{(#2#1#3)}{ and~(#2#1#3)}{, (#2#1#3)}{ and~(#2#1#3)}

\crefformat{figure}{Figure~#2#1#3}
\crefmultiformat{figure}{Figures~ #2#1#3}{ and~#2#1#3}{, (#2#1#3)}{ and~(#2#1#3)}
\crefformat{table}{Table~#2#1#3}

\def\e{\mbox{\boldmath $e$}}
\def\f{\mbox{\boldmath $f$}}

\def\g{\mbox{\boldmath $g$}}
\def\h{\mbox{\boldmath $h$}}
\def\m{\mbox{\boldmath $m$}}

\def\x{\mbox{\boldmath $x$}}
\def\y{\mbox{\boldmath $y$}}

\def\0{\mbox{\boldmath $0$}}

\begin{document}

\title[A third-order method for LLG equation]{Unstability of numerical method for Micromagnetics Simulations with large damping parameters}


\author[C. Xie]{Changjian Xie}
\address{School of Mathematics and Physics\\ Xi'an-Jiaotong-Liverpool University\\Re'ai Rd. 111, Suzhou, 215123, Jiangsu\\ China.}
\email{Changjian.Xie@xjtlu.edu.cn}

\author[C. Wang]{Cheng Wang}
\address{Mathematics Department\\ University of Massachusetts\\ North Dartmouth\\ MA 02747\\ USA.}
\email{cwang1@umassd.edu}


\date{\today}


\begin{abstract}
We propose and implement a third-order accurate numerical scheme for the Landau-Lifshitz-Gilbert equation, which describes magnetization dynamics in ferromagnetic materials under large damping parameters. This method offers two key advantages: (1) It solves only constant-coefficient linear systems, enabling fast solvers and thus achieving much higher numerical efficiency than existing second-order methods. (2) It attains third-order temporal accuracy and fourth-order spatial accuracy, and is unconditionally stable for large damping parameters. Numerical examples in 1D and 3D simulations verify both its third-order accuracy and efficiency gains. However, when large damping parameters and pre-projection solutions are involved, both this proposed method and a second-order method of the same style fail to capture reasonable physical structures, despite extensive theoretical analyses. Additionally, comparisons of domain wall dynamics among BDF2, BDF3, and BDF1 show that BDF2 and BDF3 yield failed simulations, while BDF1 performs marginally better.
\end{abstract}

\maketitle

\section{Introduction}

Ferromagnetic materials find widespread use in data storage, leveraging the bistable states of their intrinsic magnetic order or magnetization. Magnetization dynamics is modeled by the Landau-Lifshitz-Gilbert (LLG) equation~\cite{Landau1935On,Gilbert:1955}, which incorporates two key dynamic terms: an energy-conservative gyromagnetic term and an energy-dissipative damping term.

The damping term is critical as it strongly influences the energy consumption and operational speed of magnetic devices. A recent experiment on a magnetic-semiconductor heterostructure~\cite{Zhang2020ExtremelyLM} has shown that the Gilbert damping constant is tunable. At the microscopic level, electron scattering, itinerant electron relaxation~\cite{Heinrich1967TheIO}, and phonon-magnon coupling~\cite{Suhl1998TheoryOT, Nan2020ElectricfieldCO} are responsible for damping, and these can be derived from electronic structure calculations \cite{TangXia2017}. For practical applications, tuning the damping parameter enables optimization of magnetodynamic properties, such as reducing switching currents and increasing the writing speed of magnetic memory devices~\cite{Wei2012MicromagneticsAR}.

While most experiments focus on small damping parameters \cite{Budhathoki2020LowGD,Lattery2018LowGD,Weber2019GilbertDO}, large damping effects are observed in \cite{GilbertKelly1955, Tanaka2014MicrowaveAssistedMR}. A large damping constant tends to shorten magnetization switching time \cite{Tanaka2014MicrowaveAssistedMR}, and extremely large damping parameters ($\sim 9$) are reported in \cite{GilbertKelly1955}.

The LLG equation is a vectorial, nonlinear system with point-wise constant magnetization length. Significant efforts have been dedicated to developing efficient and stable numerical methods for micromagnetics simulations; see~\cite{kruzik2006recent,cimrak2007survey} for reviews. Semi-implicit schemes are popular as they avoid complex nonlinear solvers while maintaining numerical stability \cite{alouges2006convergence, gao2014optimal, Xie2018}. Specifically, a second-order backward differentiation formula (BDF) scheme with one-sided interpolation is constructed in \cite{Xie2018}, requiring the solution of a 3D linear system with non-constant coefficients at each time step. Moreover, a theoretical analysis of second-order convergence for this BDF2 method is established in \cite{jingrun2019analysis}. Alternatively, a linearly implicit method in \cite{alouges2006convergence} employs tangent space to handle the magnetization length constraint, achieving first-order temporal accuracy. High-order BDF schemes are constructed and analyzed in a more recent work~\cite{Lubich2021}. The unconditional unique solvability of semi-implicit schemes is proven in \cite{jingrun2019analysis,Lubich2021}, but their convergence analysis necessitates the temporal step-size to be proportional to the spatial grid-size. A notable drawback of these semi-implicit schemes is that the vectorial structure of the LLG equation leads to non-symmetric linear systems at each time step, which cannot be solved by FFT-based fast solvers. Instead, GMRES is often employed, whose efficiency depends heavily on the temporal step-size and spatial grid-size, and extensive numerical experiments indicate it is far more computationally expensive than standard Poisson solvers~\cite{Xie2018}.

In this paper, we propose a high-order accurate numerical method with third-order temporal accuracy and fourth-order spatial accuracy for solving the LLG equation with large damping parameters, and its complexity is comparable to solving the scalar heat equation. To achieve this, the LLG system is reformulated, with the damping term rewritten as a harmonic mapping flow. The constant-coefficient Laplacian part is treated via standard BDF3 temporal discretization, and the associated dissipation forms the basis of numerical stability. Meanwhile, all nonlinear parts, including the gyromagnetic term and the remaining nonlinear expansions in the damping term, are approximated fully explicitly using a third-order extrapolation formula. Owing to this fully explicit treatment of nonlinear parts, the resulting numerical scheme only requires a standard Poisson solver at each time step. This greatly facilitates computations, as FFT-based fast solvers can be efficiently applied due to the SPD structure of the involved linear system. Additionally, numerical stability is demonstrated through extensive computational experiments. These experiments verify that pre-projected solutions introduce significant instability, even though the dissipation property of the heat equation part can ensure the numerical stability of nonlinear parts to some extent under large damping parameters. As the order of the numerical method increases, numerical simulations become more unstable, and BDF1 results outperform those of BDF2 and BDF3. This is because higher-order methods rely more on solution information from previous unprojected steps, exerting a greater impact on stability—especially in stray field updates.

The remainder of this paper is organized as follows. In \cref{sec: numerical scheme}, the micromagnetics model is reviewed, the numerical method is proposed, and comparisons with BDF2 and BDF1 methods are provided. Subsequently, numerical results are presented in \cref{sec:experiments}, including checks of temporal and spatial accuracy in 1D and 3D computations, investigations of numerical efficiency (compared to BDF2 and BDF1 algorithms), studies on stability with respect to the damping parameter, and analyses of domain wall motion instability. Finally, concluding remarks are given in \cref{sec:conclusions}.

\section{The physical model and the numerical method}
\label{sec: numerical scheme}

\subsection{Landau-Lifshitz-Gilbert equation}

The LLG equation describes the dynamics of magnetization which consists of the gyromagnetic term and the damping term~\cite{Landau1935On,Brown1963micromagnetics}. In the nondimensionalized form, this equation reads as
\begin{align}\label{c1-large}
{\m}_t =-{\m}\times{\bm h}_{\text{eff}}-\alpha{\m}\times({\m}\times{\bm h}_{\text{eff}})
\end{align}
with the homogeneous Neumann boundary condition
\begin{equation}\label{boundary-large}
\frac{\partial{\m}}{\partial {\bm \nu}}\Big|_{\partial \Omega}=0,
\end{equation}
where $\Omega$ is a bounded domain occupied by the ferromagnetic material and $\bm \nu$ is unit outward normal vector along $\partial \Omega$. 

In more details, the magnetization ${\m}\,:\,\Omega\subset\mathbb{R}^d\to \mathbb{R}^3,d=1,2,3 $ is a three-dimensional vector field with a pointwise constraint $|\m|=1$. The first term on the right-hand side in \cref{c1-large} is the gyromagnetic term and the second term stands for the damping term, with $\alpha>0$ being the dimensionless damping coefficient.

The effective field ${\bm h}_{\text{eff}}$ is obtained by taking the variation of the Gibbs free energy of the magnetic body with respect to $\m$. The free energy includes the exchange energy, the anisotropy energy, the magnetostatic energy, and the Zeeman energy:  
\begin{equation}\label{LL-Energy}
F[\m] = \frac {\mu_0 M_s^2}{2} \left\{\int_\Omega \left( \epsilon|\nabla\m|^2 +
q\left(m_2^2 + m_3^2\right)
-2\h_e\cdot\m - \h_s\cdot\m \right)\mathrm{d}\x \right\} . 
\end{equation}
Therefore, the effective field includes the exchange field, the anisotropy field, the stray field $\h_s$, and the external field $\h_e$. For a uniaxial material, it is clear that 
\begin{align}
{\bm h}_{\text{eff}} =\epsilon\Delta\m-q(m_2\e_2+m_3\e_3)+\h_s+\h_e,
\end{align}
where the dimensionless parameters become $\epsilon=C_{ex}/(\mu_0 M_s^2L^2)$ and $q=K_u/(\mu_0 M_s^2)$ with $L$ the diameter of the ferromagnetic body and $\mu_0$ the permeability of vacuum. The unit vectors are given by ${\bm e}_2=(0,1,0)$, ${\bm e}_3=(0,0,1)$, and $\Delta$ denotes the standard Laplacian operator. For the Permalloy, an alloy of Nickel ($80\%$) and Iron ($20\%$), typical values of the physical parameters are given by: the exchange constant $C_{ex}=1.3\times 10^{-11}\,\textrm{J/m}$, the anisotropy constant $K_u = 100\, \textrm{J/}\textrm{m}^3$, the saturation magnetization constant $M_s = 8.0\times 10^{5}\,\textrm{A/m}$. The stray field takes the form
\begin{align}\label{eqn:div}
	{\h}_{\text{s}}=\frac{1}{4\pi}\nabla \int_{\Omega} \nabla\left( \frac{1}{|\x-\y|}\right)\cdot {\bm m}({\bm y})\,d{\bm y}.
\end{align}
If $\Omega$ is a rectangular domain, the evaluation of \eqref{eqn:div} can be efficiently done by the Fast Fourier Transform (FFT) \cite{Wang2000}.

For brevity, the following source term is defined 
\begin{align}\label{eq-4}
\f=-Q(m_2\e_2+m_3\e_3)+\h_s+\h_e.
\end{align}
and the original PDE system \cref{c1-large} could be rewritten as
\begin{align}\label{eq-5}
\m_t=-\m\times(\epsilon\Delta\m+\f)-\alpha\m\times\m\times(\epsilon\Delta\m+\f).
\end{align}
Thanks to point-wise identity $|\m|=1$, we obtain an equivalent form: 
\begin{equation}\label{eq-model}
\m_t=\alpha  (\epsilon\Delta\m+\f)+\alpha \left(\epsilon |\nabla \m|^2 -\m \cdot\f \right)\m-\m\times(\epsilon\Delta\m+\f).
\end{equation}
In particular, it is noticed that the damping term is rewritten as a harmonic mapping flow, which contains a constant-coefficient Laplacian diffusion term. This fact will greatly improve the numerical stability of the proposed scheme.  

For the numerical description, we first introduce some notations for discretization and numerical approximation. 
Denote the temporal step-size by $k$, and $t^n=nk$, $n\leq \left\lfloor\frac{T}{k}\right\rfloor$ with $T$ the final time. The spatial mesh-size is given by $h_x=h_y=h_z=h=1/N$, and $\m_{i,j,\ell}^n$ stands for the magnetization at time step $t^n$, evaluated at the spatial location $(x_{i-\frac12},y_{j-\frac12},z_{\ell-\frac12})$ with $x_{i-\frac12}=\left(i-\frac12\right)h_x$, $y_{j-\frac12}=\left(j-\frac12\right)h_y$ and $z_{\ell-\frac12}=\left(\ell-\frac12\right)h_z$ ($0\leq i,j,\ell\leq N+1$). In addition, a third order extrapolation formula is used to approximate the homogeneous Neumann boundary condition. For example, such a formula near the boundary along the $z$ direction is given by 
\begin{align*}
\m_{i,j,-1}=\m_{i,j,2},\quad \m_{i,j,1}=\m_{i,j,0},\quad \m_{i,j,N+1}=\m_{i,j,N}\quad \m_{i,j,N+2}=\m_{i,j,N-1}.
\end{align*}
The boundary extrapolation along other boundary sections can be similarly made.

The standard second-order centered difference applied to $\Delta \m$ results in
\begin{align*}
\Delta_h \m_{i,j,k} &=\frac{\delta_x^2\m_{i,j,k}}{h_x^2}
+\frac{\delta_y^2\m_{i,j,k}}{h_y^2}
+\frac{\delta_z^2\m_{i,j,k}}{h_z^2},
\end{align*}
where $\delta_x^2 \m_{i,j,k}=\m_{i+1,j,k}-2\m_{i,j,k}+\m_{i-1,j,k}$, $\delta_y^2 \m_{i,j,k}=\m_{i,j+1,k}-2\m_{i,j,k}+\m_{i,j-1,k}$ and $\delta_z^2 \m_{i,j,k}=\m_{i,j,k+1}-2\m_{i,j,k}+\m_{i,j,k-1}$
and the discrete gradient operator $\nabla_h \m$ with $\m=(u, v, w)^T$ reads as
\begin{align*}
	\nabla_h\m_{i,j,k} =\left(\frac{\delta_x \m_{i,j,k}}{h_x},\frac{\delta_y \m_{i,j,k}}{h_y},\frac{\delta_z \m_{i,j,k}}{h_z}\right)^T,
\end{align*}
where $\delta_x \m_{i,j,k}=\m_{i+1,j,k}-\m_{i-1,j,k}$, $\delta_y \m_{i,j,k}=\m_{i,j+1,k}-\m_{i,j-1,k}$ and $\delta_z \m_{i,j,k}=\m_{i,j,k+1}-\m_{i,j,k-1}$.

Subsequently, the BDF1 and the BDF2 numerical methods need to be reviewed, which could be used for the later comparison. 

\subsection{The first-order BDF method}

The first-order BDF (BDF1) method is based on a Backward Differentiation Formula, combined with an explicit extrapolation. The cross product nonlinear term is treated with previous step solution. It only requires a linear equation solvers with constant coefficients; as a result, the FFT-based fast solvers could be easily applied. This method is first-order in time and second-order in space. Below is the detailed outline of the BDF1 method.

\begin{align*}
\left\{\begin{aligned}
&\frac{\tilde{\m}_h^{n+1}-\tilde{\m}_h^{n}}{k}=-{\m}_h^{n} \times (\epsilon \Delta_{h}{\tilde{\m}}_h^{n}+{{\f}}_h^{n})+\alpha(\epsilon\Delta_{h}\tilde{\m}_h^{n+1}+{\f}_h^{n})\\
&\qquad\qquad+\alpha(\epsilon|\tilde{\nabla}_{h}{\m}_h^{n}|^2-{\m}_h^{n}\cdot {\f}_h^{n}) {\m}_h^{n},\\
&\m_h^{n+1}=\frac{\tilde{\m}_h^{n+1}}{|\tilde{\m}_h^{n+1}|},
\end{aligned}
\right.
\end{align*}

\subsection{The second-order BDF method}
Such approach has been outlined in \cite{CaiChenWangXie2022simulation} and \cite{CaiChenWangXie2023analysis}. This method is based on the second-order BDF temporal discretization, combined with an explicit extrapolation. It is found that BDF2 is unconditionally stable and is second-order accurate in both space and time. The algorithmic details are given as follows.
\begin{equation}\label{sipm}
\left\{ 
\begin{aligned}
&\frac{\frac32 {\tilde{\m}}_h^{n+2} - 2 \tilde{\m}_h^{n+1} + \frac12 \tilde{\m}_h^n}{k}
=  - \hat{\m}_h^{n+2} \times\big(\epsilon \Delta_h\hat{\tilde{\m}}_h^{n+2} +\hat{\f}_h^{n+2} \big) \\
&\quad +\alpha(\epsilon\Delta_h\tilde{\m}_h^{n+2}+\hat{\f}_h^{n+2})+\alpha\left(\epsilon|\tilde{\nabla}_h\hat{\m}_h^{n+2}|^2-\hat{\m}_h^{n+2}\cdot\hat{\f}_h^{n+2}\right)\hat{\m}_h^{n+2}, \\
& \qquad\qquad\qquad\qquad\quad \m_h^{n+2} = \frac{\tilde{\m}_h^{n+2}}{ |\tilde{\m}_h^{n+2}| },
\end{aligned}
\right.
\end{equation} 
where $\tilde{\m}_h^{n+2}$ is an intermediate magnetization, and $\hat{\m}_h^{n+2}$, $\hat{\f}_h^{n+2}$ are given by the following extrapolation formula: 
\begin{align*}
\hat{\m}_h^{n+2} &=2{\m}_h^{n+1}-{\m}_h^n, \label{m_hat}\\
\hat{\f}_h^{n+2} &=2{\f}_h^{n+1}-{\f}_h^n,
\end{align*}
with $\f_h^{n}=-Q(m_2^n\e_2+m_3^n\e_3)+\h_s^n+\h_e^n$.

\subsection{The proposed third method} \label{discretisations}

The high-order BDF idea leads to the proposed numerical method as follows. 
\begin{equation}\label{proposed}
\left\{ 
\begin{aligned}
&\frac{\frac{11}{6} \tilde{\m}_h^{n+3} - 3 \tilde{\m}_h^{n+2} +\frac32  \tilde{\m}_h^{n+1}-\frac13 \tilde{\m}_h^n}{k}
=  - \hat{\m}_h^{n+3} \times \left(\epsilon \Delta_h \hat{\tilde{\m}}_h^{n+3} +\hat{\f}_h^{n+3}\right) \\
&\quad + \alpha \left(\epsilon \Delta_h \tilde{\m}_h^{n+3}+\hat{\f}_h^{n+3}\right)\\  
&\quad + \alpha \left(\epsilon | \tilde{\nabla}_{h,(4)} \hat{\m}_h^{n+3} |^2-\hat{\m}_h^{n+3}\cdot \hat{\f}_h^{n+3}\right) \hat{\m}_h^{n+3},\\ 
&\m_h^{n+3} = \frac{\tilde{\m}_h^{n+3}}{ |\tilde{\m}_h^{n+3}| } ,
\end{aligned}
\right.
\end{equation}
where
\begin{align*}
\hat{\m}_h^{n+3} &= 3 \m_h^{n+2} - 3\m_h^{n+1}+\m_h^n,\\
\hat{\f}_h^{n+3} &= 3 \f_h^{n+1} -3 \f_h^{n+1}+\f_h^n.
\end{align*}

\cref{tab-features} compares the proposed method, the BDF2 and the BDF1 in terms of number of unknowns, dimensional size, symmetry pattern, and availability of FFT-based fast solver of linear systems of equations, and the number of stray field updates. At the formal level, the proposed method is clearly superior to both the BDF2 and the BDF1 algorithms in terms of accuracy. In more details, this scheme will greatly improve the computational efficiency, since only three Poisson solvers are needed at each time step. Moreover, this numerical method preserves a third-order accuracy in time and fourth-order accuracy in space. Interestingly, the numerical results in \cref{sec:experiments} will demonstrate that the proposed scheme provides a subtle approach for micromagnetics simulations with less stability in the regime of large damping parameters. 
\begin{table}[htbp]
	\begin{center}
		\caption{Comparison of the proposed method, the first-order method, and the second-order method.}\label{tab-features}
		\begin{tabular}{cccc}
			\hline
			Property or number & Proposed method & BDF2 & BDF1\\
			\hline
			Linear systems& \boldsymbol{$3$} & $3$ & $3$ \\
			Size & \boldsymbol{$N^3$} & $N^3$& $N^3$ \\
			Symmetry& {\bf Yes}& Yes& Yes \\
			Fast Solver& {\bf Yes}& Yes& Yes \\
			Accuracy& \boldsymbol{$\mathcal{O}(k^3+h^4)$} & $\mathcal{O}(k^2+h^2)$ & $\mathcal{O}(k+h^2)$ \\
			Stray field updates & \boldsymbol{$1$} &$1$ &$1$ \\
			\hline
		\end{tabular}
	\end{center}
\end{table}

\begin{remark}
To kick start the proposed method, one can apply a first-order and a second-order algorithm, such as the first-order BDF method and the second-order BDF method, in the first and second time step. 
An overall third-order accuracy is preserved in such an approach.
\end{remark}

\section{Numerical experiments}
\label{sec:experiments}

In this section, we present a few numerical experiments with a sequence of damping parameters for the proposed method, the BDF2 introduced by \cite{CaiChenWangXie2022simulation} and \cite{CaiChenWangXie2023analysis} and the BDF1, with the accuracy, efficiency, and stability examined in details. Domain wall dynamics is studied and its velocity is recorded in terms of the damping parameter and the external magnetic field. 

\subsection{Accuracy and efficiency tests}

We set $\epsilon=1$ and $\f=0$ in \cref{eq-model} for convenience. The 1D exact solution is given by 
\begin{equation*}
	\m_e=\left(\cos(\cos(\pi x))\sin t, \sin(\cos(\pi x))\sin t, \cos t\right)^T,
\end{equation*}
and the corresponding exact solution in 3D becomes 
\begin{equation*}
\m_e=\left(\cos(\cos(\pi x)\cos(\pi y)\cos(\pi z))\sin t, \sin(\cos (\pi x)\cos(\pi y)\cos(\pi z))\sin t, \cos t\right)^T,
\end{equation*}
In fact, the above exact solutions satisfy \cref{eq-model} with the forcing term $\g=\partial_t \m_e-\alpha \Delta \m_e -\alpha |\nabla \m_e|^2+\m_e \times \Delta \m_e$, as well as the homogeneous Neumann boundary condition. 

For the temporal accuracy test in the 1D case, we fix the spatial resolution as $h=1D-4$, so that the spatial approximation error becomes negligible. The damping parameter is taken as $\alpha=10$, and the final time is set as $T=0.1$. In the 3D test for the temporal accuracy, 
due to the limitation of spatial resolution, we take a sequence of spatial and temporal mesh sizes: $k=h_x^2=h_y^2=h_z^2=h^2=T/N_0$ for the first-order method and $k=h_x=h_y=h_z=h=T/N_0$ for the second-order method, and $k=h_x^{\frac43}=h_y^{\frac43}=h_z^{\frac43}=h^{\frac43}=T/N_0$ for the proposed method, with the variation of $N_0$ indicated below. Similarly, the damping parameter is given by $\alpha=10$, while the final time $T$ is indicated below. In turn, the numerical errors are recorded in term of the temporal step-size $k$ in \cref{tab-1}. It is clear that the temporal accuracy orders of the proposed numerical method, the BDF2, and the BDF1 are given by $3$, $2$, and $1$, respectively, in both the 1D and 3D computations. 

\begin{table}[htbp] 
	\centering
	{\caption{The numerical errors for the proposed method, the BDF1 and the BDF2 with $\alpha=10$ and $T=0.1$. Left: 1D with $h=1D-4$; Right: 3D with $k=h_x^2=h_y^2=h_z^2=h^2=T/N_0$ for BDF1 and $k=h_x=h_y=h_z=h=T/N_0$ for the BDF2, and $k=h_x^{\frac43}=h_y^{\frac43}=h_z^{\frac43}=h^{\frac43}=T/N_0$ for the proposed method with $N_0$ specified in the table.}\label{tab-1} }{
		\subfloat[Proposed method ]{\label{tab:floatrow:one}%
			\begin{tabular}{cccc|cccc} 
				\hline	
				1D  & & {} & {} &3D &{} & {} &{} \\
				$k$ & $\|\cdot\|_{\infty}$ & $\|\cdot\|_{2}$ & $\|\cdot\|_{H^1}$ & 	$k,k^3\approx h^4$ & $\|\cdot\|_{\infty}$ & $\|\cdot\|_{2}$ & $\|\cdot\|_{H^1}$ \\
				\hline
			$T/8$& 1.982D-7& 1.388D-7 &6.165D-7 & $T/4$ & 8.588D-6 & 1.754D-6 & 3.559D-5 \\	
			$T/12$ & 5.829D-8 & 4.227D-8& 1.881D-7&	$T/5$ &  3.996D-6& 8.276D-7& 1.837D-5\\
			$T/16$ &  2.485D-8&1.712D-8 & 7.577D-8&	 $T/6$ & 2.308D-6 & 4.752D-7 & 1.011D-5\\
			$T/24$ & 7.529D-9 &4.910D-9 & 2.141D-8 &	 $T/8$ & 1.026D-6 & 2.078D-7 & 4.492D-6\\
			$T/32$ & 3.027D-9& 2.286D-9 &1.019D-8  & 	$T/9$& 7.723D-7 & 1.566D-7 & 3.369D-6 \\
				order & 3.00 & 2.99 & 3.00 &{--}& 2.96&2.98 & 2.93\\
				
				
				%
				\hline
			\end{tabular}	
		}
		\qquad
		\subfloat[BDF1]{\label{tab:floatrow:two}
			\begin{tabular}{cccc|cccc} 
				\hline
				1D & &  & {} & 3D & {} &  & {} \\
				$k$ & $\|\cdot\|_{\infty}$ & $\|\cdot\|_{2}$ & $\|\cdot\|_{H^1}$ & $k=h^2$ & $\|\cdot\|_{\infty}$ & $\|\cdot\|_{2}$ & $\|\cdot\|_{H^1}$\\
				\hline
				$T/8$ & 1.346D-3& 9.594D-4 & 4.258D-3& $T/40$ &  5.443D-4& 9.439D-5& 5.458D-4 \\
				$T/12$ & 9.232D-4 & 6.525D-4 & 2.882D-3 & $T/57$ & 3.809D-4 & 6.581D-5 & 3.786D-4 \\
				$T/16$ & 7.095D-4 & 4.988D-4 & 2.192D-3 & $T/78$ & 2.796D-4 & 4.821D-5 & 2.769D-4  \\
				$T/24$ &  4.918D-4 & 3.441D-4 & 1.498D-3  &$T/102$& 2.142D-4 & 3.688D-5 & 2.115D-4\\
				$T/32$ &  3.799D-4 & 2.655D-4 & 1.147D-3&$T/129$ & 1.694D-4 & 2.914D-5 & 1.669D-4 \\
				order & 0.91 & 0.93 &0.95 &{--}&1.00 &1.00&1.01 \\
				
				
				%
				\hline
			\end{tabular}
		}
		\qquad
		\subfloat[BDF2]{\label{tab:floatrow:three}
			\begin{tabular}{cccc|cccc} 
				\hline
				1D  & & & {} & 3D & {} &  & {}\\
				$k$ & $\|\cdot\|_{\infty}$ & $\|\cdot\|_{2}$ & $\|\cdot\|_{H^1}$ & $k=h$ & $\|\cdot\|_{\infty}$ & $\|\cdot\|_{2}$ & $\|\cdot\|_{H^1}$\\
				\hline
				$T/8$& 5.258D-6& 3.843D-6& 1.028D-5 & $T/3$ & 1.639D-4 & 2.832D-5& 2.263D-4\\	
			$T/12$&  2.472D-6& 1.816D-6& 4.748D-6 & $T/4$ & 9.423D-5 & 1.619D-5 & 1.296D-4\\
				$T/16$ & 1.428D-6& 1.052D-6& 2.721D-6  &	$T/5$ & 6.148D-5& 1.060D-5& 8.264D-5 \\
			$T/24$ & 6.518D-7& 4.811D-7& 1.232D-6 &	 $T/6$ & 4.319D-5 & 7.485D-6 & 5.742D-5\\
				$T/32$ & 3.716D-7& 2.744D-7& 7.000D-7  & $T/7$& 3.200D-5 & 5.566D-6 & 4.223D-5 \\
				order &1.91 & 1.91&  1.94& {--}& 1.93&1.92 &1.98\\
				
				
				%
				\hline
			\end{tabular}
	} }	
\end{table}

The spatial accuracy order is tested by fixing $k=1D-5$, $\alpha=10$, $T=0.1$ in 1D and $k=1D-4$, $\alpha=10$, $T=0.1$ in 3D. The numerical error is recorded in term of the spatial grid-size $h$ in \cref{tab-2}. Similarly, the presented results have indicated the second order spatial accuracy of the BDF2 and BDF1 algorithms, and the fourth-order spatial accuracy for the proposed method, in both the 1D and 3D computations. 

\begin{table}[htbp]
	\centering
	{\caption{The numerical errors of the proposed method, the BDF1 and the BDF2 with $\alpha=10$ and $T=0.1$. Left: 1D with $k=1D-5$; Right: 3D with $k=1D-4$.} \label{tab-2} }{
		\subfloat[Proposed method ]{\label{tab:floatrow:1-S}
			\begin{tabular}{cccc|cccc}	
				\hline
				1D  & & & {} & 3D & & & \\
				$h$ & $\|\cdot\|_{\infty}$ &$\|\cdot\|_{2}$ &$\|\cdot\|_{H^1}$& $h$ & $\|\cdot\|_{\infty}$ & $\|\cdot\|_{2}$ & $\|\cdot\|_{H^1}$  \\
				\hline
				1/16 &7.726D-6 &5.837D-6 &9.863D-5 & 1/16 &7.042D-6 & 1.732D-6 & 3.518D-5 \\
				1/32 &5.044D-7 &3.708D-7&6.358D-6 & 1/20 &3.066D-6 & 7.398D-7 & 1.425D-5\\
				1/64 &3.188D-8 &2.328D-8& 4.006D-7& 1/24 &1.474D-6 & 3.717D-7 & 6.792D-6\\
				1/128 &1.999D-9 &1.456D-9&2.509D-8 & 1/28 & 8.031D-7 & 2.043D-7 & 3.662D-6\\
				1/256 &1.248D-10 &9.110D-11&1.569D-9 & 1/32 &4.734D-7& 1.232D-7& 2.141D-6\\
				order  &3.98 &3.99& 3.99& {--} & 3.91&3.82& 4.04 \\
				\hline
			\end{tabular}
		}	
		\qquad
		\subfloat[BDF1]{\label{tab:floatrow:2-S}
			\begin{tabular}{cccc|cccc}	
				\hline
				1D  & &  & {} & 3D & & & \\
				$h$ & $\|\cdot\|_{\infty}$ &$\|\cdot\|_{2}$ &$\|\cdot\|_{H^1}$ & $h$ & $\|\cdot\|_{\infty}$ &$\|\cdot\|_{2}$ &$\|\cdot\|_{H^1}$\\
				\hline
				1/16 & 3.048D-4 & 2.818D-4 & 2.011D-3  & 1/16 &4.577D-4& 9.320D-5& 7.925D-4\\
				1/32 & 7.700D-5& 7.017D-5 & 5.015D-4 & 1/20 & 2.952D-4& 5.960D-5& 5.035D-4 \\
				1/64 & 1.927D-5& 1.756D-5& 1.252D-4 & 1/24 & 2.061D-4 & 4.139D-5 & 3.479D-4 \\
				1/128 & 4.794D-6& 4.429D-6 & 3.120D-5 & 1/28 &1.521D-4 & 3.042D-5& 2.547D-4 \\
				1/256 & 1.183D-6& 1.150D-6& 7.714D-6 & 1/32 &1.170D-4 & 2.329D-5 & 1.945D-4 \\
				order  &2.00 &1.99&2.01 & {--} &1.97 &2.00&2.03 \\
				\hline
			\end{tabular}
		}
		\qquad
		\subfloat[BDF2]{\label{tab:floatrow:3-S}
			\begin{tabular}{cccc|cccc}	
				\hline
				1D  & &  & & 3D & & & \\
				$h$ & $\|\cdot\|_{\infty}$ &$\|\cdot\|_{2}$ &$\|\cdot\|_{H^1}$ &$h$ & $\|\cdot\|_{\infty}$ &$\|\cdot\|_{2}$ &$\|\cdot\|_{H^1}$ \\
				\hline
				1/16 &3.145D-4&2.939D-4&1.982D-3 & 1/16 & 4.568D-4 & 9.311D-5 & 7.928D-4\\
				1/32 &7.947D-5&7.316D-5&4.941D-4 & 1/20 & 2.943D-4 & 5.950D-5 & 5.039D-4 \\
				1/64 &1.992D-5 &1.827D-5&1.235D-4 & 1/24 & 2.051D-4 & 4.130D-5 & 3.483D-4\\
				1/128 &4.982D-6&4.567D-6& 3.086D-5& 	1/28 & 1.510D-4 & 3.033D-5 & 2.551D-4\\
				1/256&1.247D-6&1.142D-6&7.714D-6 & 	1/32 &1.158D-4 & 2.322D-5 & 1.948D-4\\
				order  &2.00&2.00&2.00& {--} &1.98& 2.00&2.02 \\
				\hline
			\end{tabular} 
		}
	}
\end{table}

\begin{figure}[htbp]
	\centering
	\subfloat[Varying $k$ in 1D up to $T=0.1$ ]{\label{cputime_1D}\includegraphics[width=2.5in]{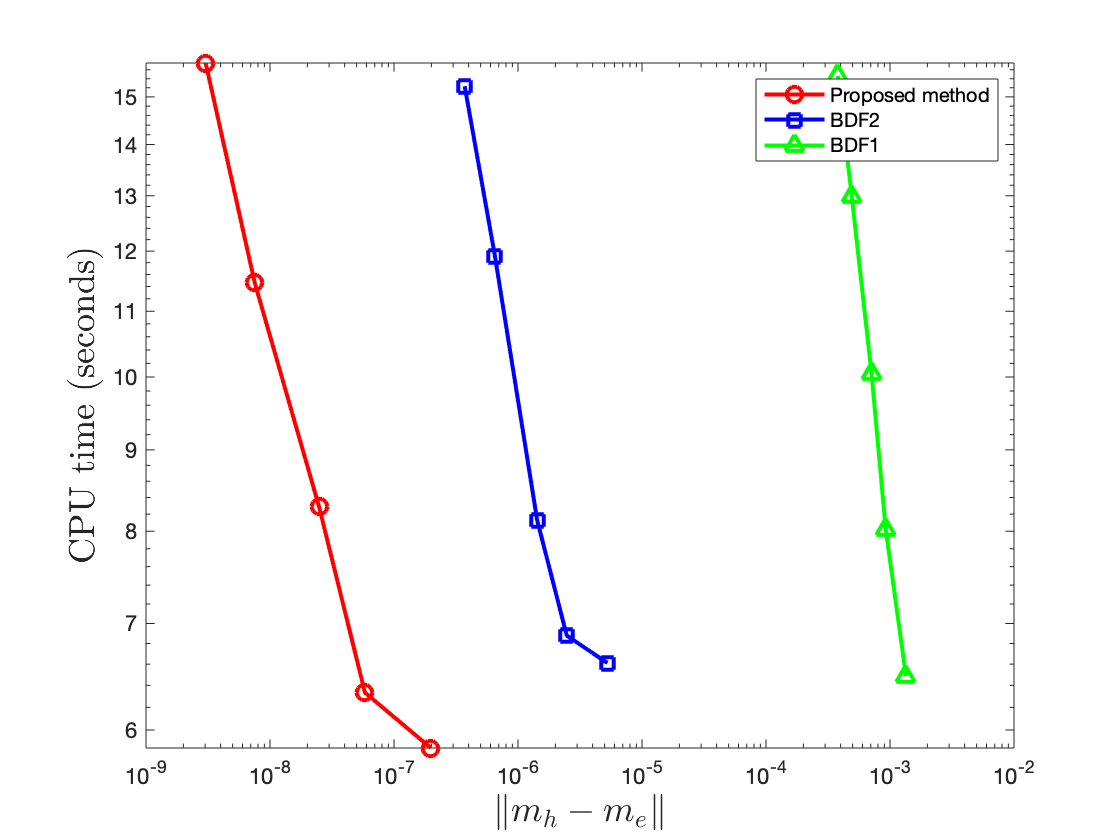}}
	\subfloat[Varying $k$ in 3D up to $T=0.1$]{\label{cputime_3D}\includegraphics[width=2.5in]{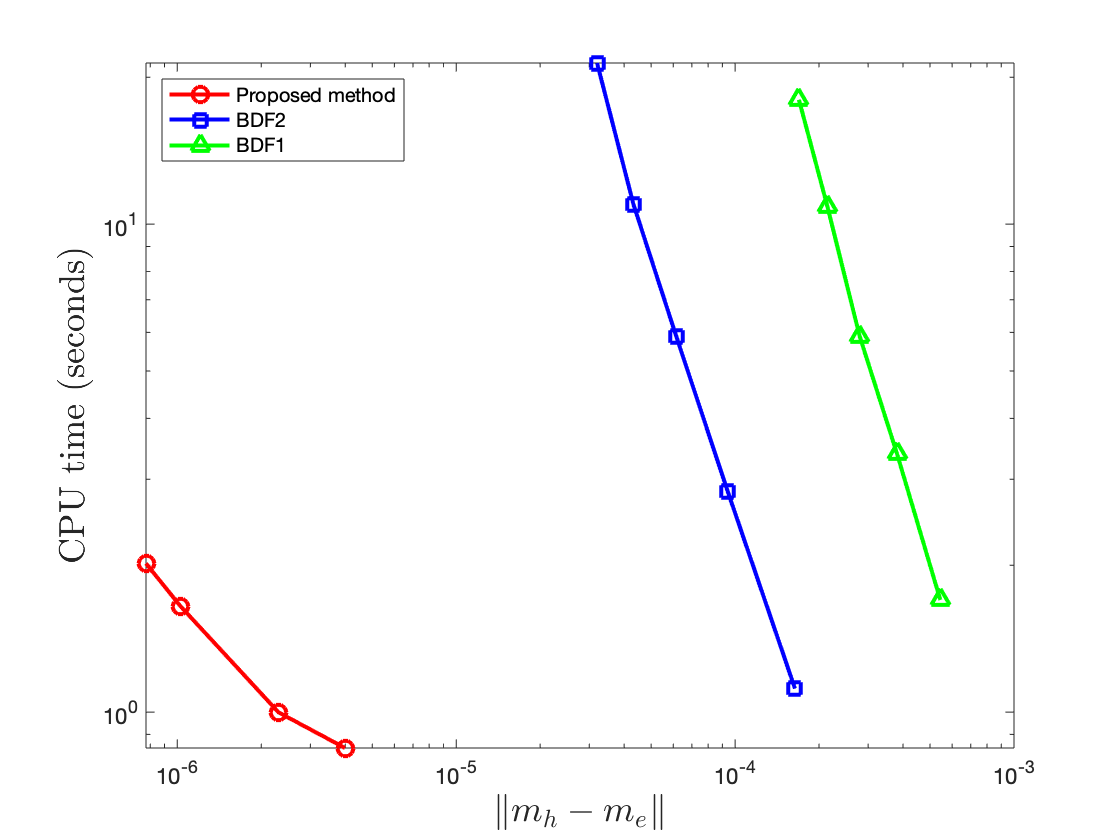}}
	\hspace{0.1in}
	\subfloat[Varying $h$ in 1D up to $T=0.1$]{\label{cputime_1D_space}\includegraphics[width=2.5in]{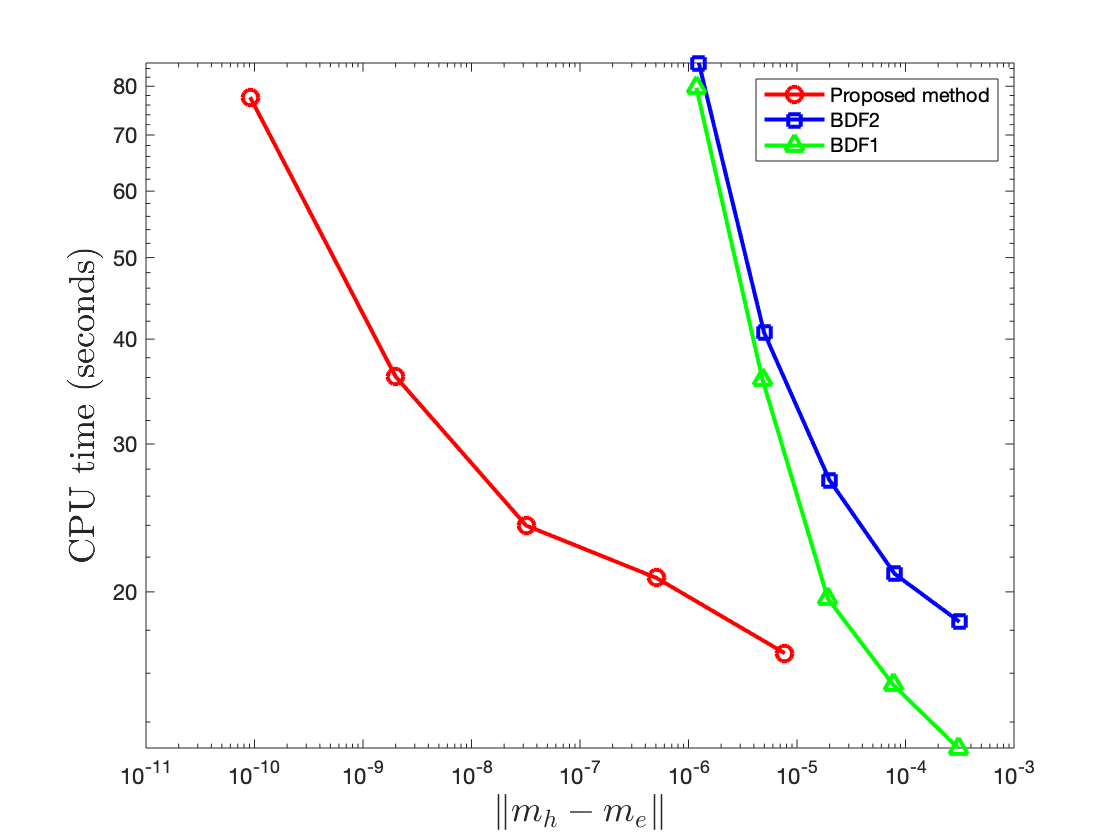}}
	\subfloat[Varying $h$ in 3D up to $T=0.1$]{\label{cputime_3D_space}\includegraphics[width=2.5in]{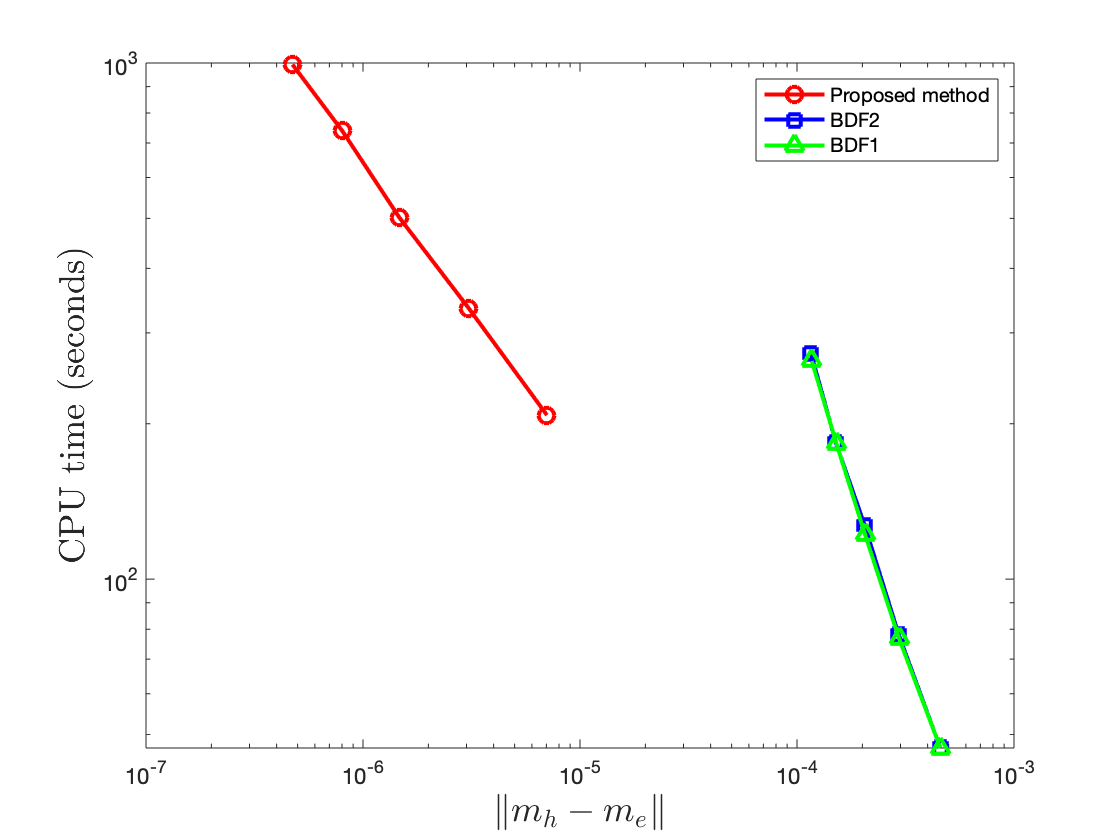}}
	\caption{CPU time needed to achieve the desired numerical accuracy, for the proposed method, the BDF2 and the BDF1 method, in both the 1D and 3D computations. The CPU time is recorded as a function of the approximation error by varying $k$ or $h$ independently. CPU time with varying $k$: proposed method $<$ BDF2 $<$ BDF1; CPU time with varying $h$: proposed method $<$ BDF1 $\lessapprox$ BDF2.}\label{cputime}
\end{figure}

To make a comparison in terms of the numerical efficiency, we plot the CPU time (in seconds) vs. the error norm $\|\m_h-\m_e\|_{\infty}$. In details, the CPU time is recorded as a function of the approximation error in \cref{cputime_1D} in 1D and in \cref{cputime_3D} in 3D, with a variation of $k$ and a fixed value of $h$. Similar plots are also displayed in \cref{cputime_1D_space} in 1D and \cref{cputime_3D_space} in 3D, with a variation of $h$ and a fixed value of $k$. In the case of a fixed spatial resolution $h$, the proposed method is significantly more efficient than the BDF2 and the BDF1 in both the 1D and 3D computations. The BDF2 is more efficient than the BDF1, while such an advantage may vary for different values of $k$ and $h$. In the case of a fixed time step size $k$, the proposed method is more efficient than the BDF2, in both the 1D and 3D computations, and the cost of BDF1 is comparable to the BDF2.  

\subsection{Stability test with large damping parameters}
\begin{figure}[htbp]
	\centering
	\subfloat{\label{BDF3_alpha_1_ang}\includegraphics[width=1.3in]{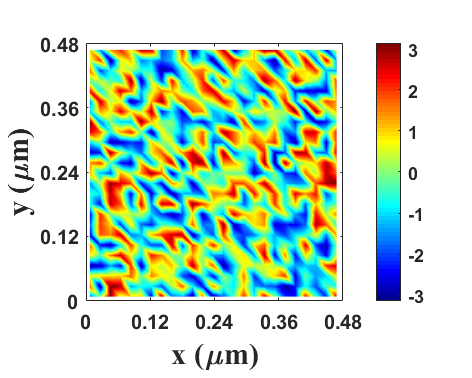}}
	\subfloat{\label{BDF3_alpha_5_ang}\includegraphics[width=1.3in]{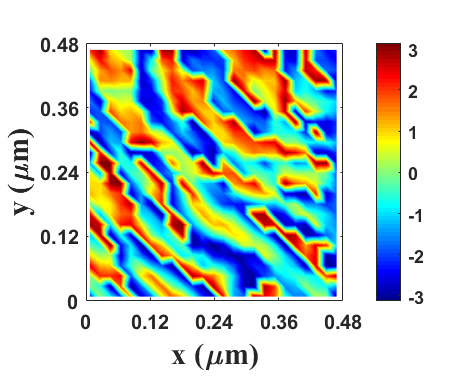}}
	\subfloat{\label{BDF3_alpha_10_ang}\includegraphics[width=1.3in]{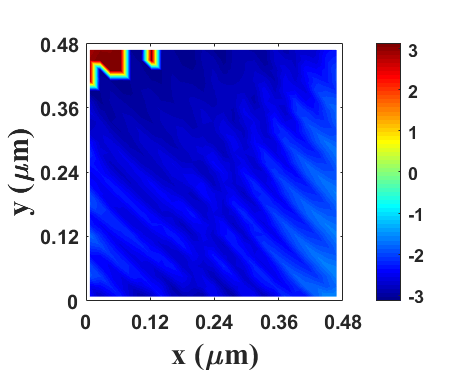}}
	\subfloat{\label{BDF3_alpha_40_ang}\includegraphics[width=1.3in]{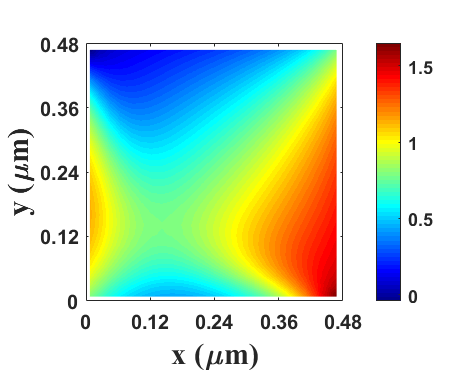}}
	\subfloat{\label{BDF3_alpha_100_ang}\includegraphics[width=1.3in]{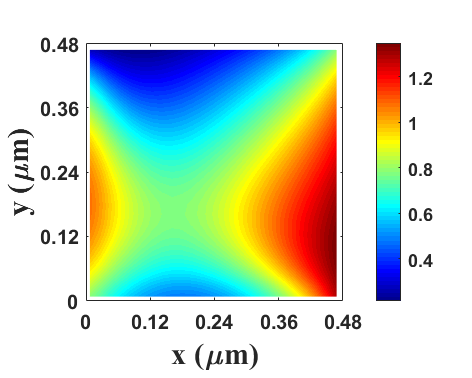}}
	\hspace{0.1in}
	\subfloat{\label{BDF2_alpha_1_ang}\includegraphics[width=1.3in]{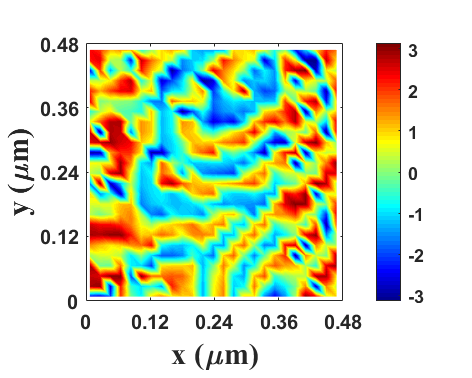}}
	\subfloat{\label{semiBDF2_alpha_5_ang}\includegraphics[width=1.3in]{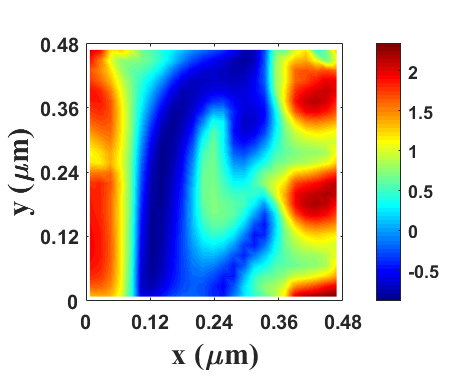}}
	\subfloat{\label{semiBDF2_alpha_10_ang}\includegraphics[width=1.3in]{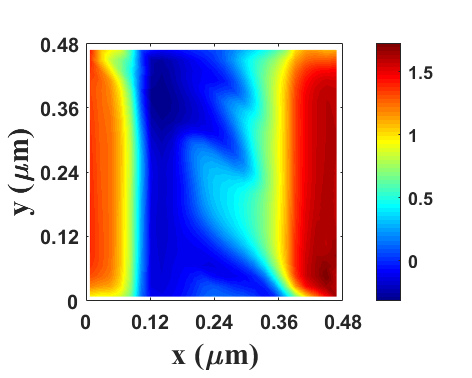}}
	\subfloat{\label{semiBDF2_alpha_40_ang}\includegraphics[width=1.3in]{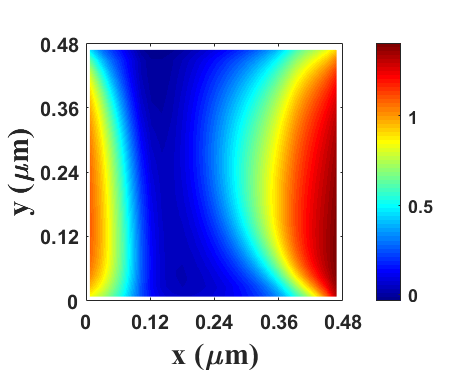}}
	\subfloat{\label{semiBDF2_alpha_100_ang}\includegraphics[width=1.3in]{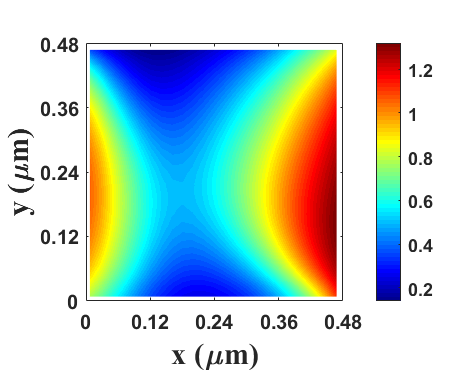}}
		\hspace{0.1in}
	\subfloat{\label{BD1_alpha_1_ang}\includegraphics[width=1.3in]{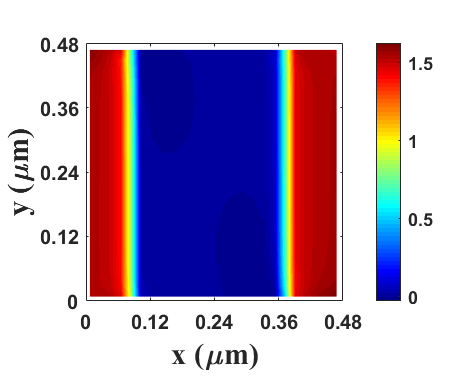}}
	\subfloat{\label{BDF1_alpha_5_ang}\includegraphics[width=1.3in]{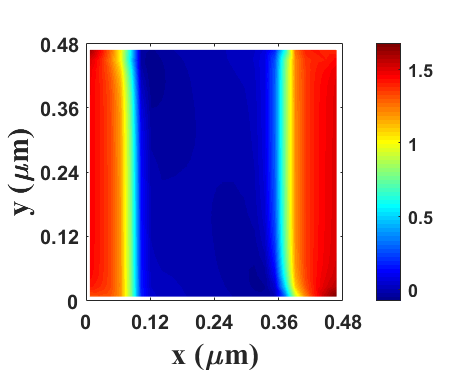}}
	\subfloat{\label{BDF1_alpha_10_ang}\includegraphics[width=1.3in]{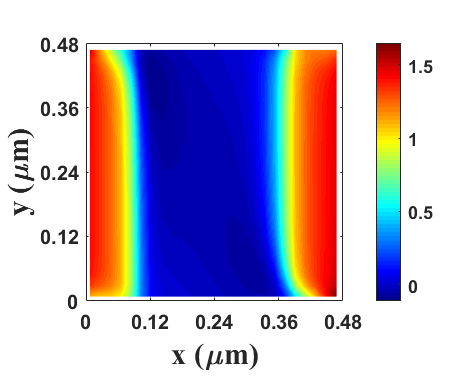}}
	\subfloat{\label{BDF1_alpha_40_ang}\includegraphics[width=1.3in]{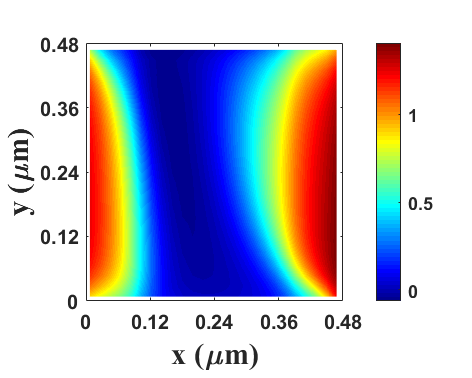}}
	\subfloat{\label{BDF1_alpha_100_ang}\includegraphics[width=1.3in]{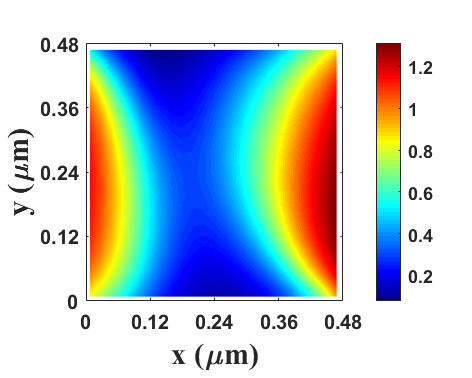}}
	\caption{Stable structures in the absence of magnetic field at $2\,$ns when $\alpha=1,5,10,40,100$. The color denotes the angle between the first two components of the magnetization vector. Top: Proposed method; Middle: BDF2; Bottom: BDF1; From left to right: $\alpha=1,5,10,40,100$, $dt=1\;ps$. }\label{BDF3_BDF2_BDF1_alpha}
\end{figure}

\begin{figure}[htbp]
	\centering
	\subfloat{\label{BDF3_alpha_1_ang_v1}\includegraphics[width=1.3in]{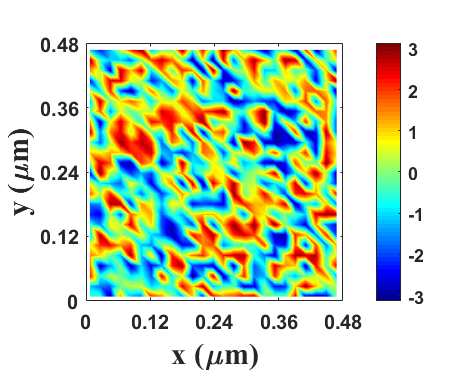}}
	\subfloat{\label{BDF3_alpha_5_ang_v1}\includegraphics[width=1.3in]{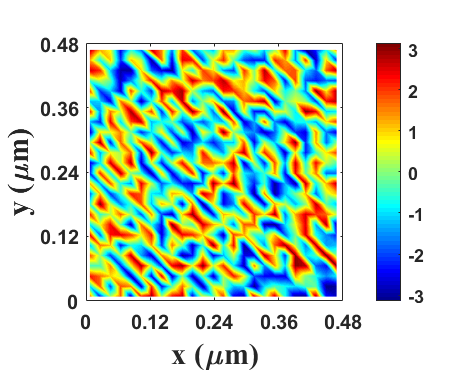}}
	\subfloat{\label{BDF3_alpha_10_ang_v1}\includegraphics[width=1.3in]{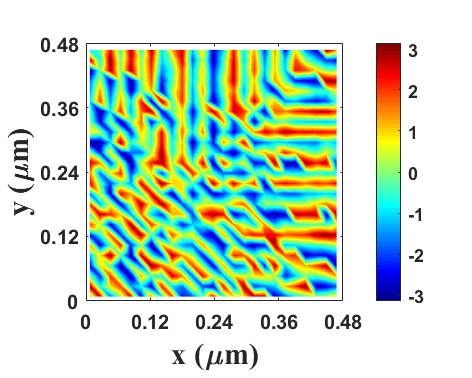}}
	\subfloat{\label{BDF3_alpha_40_ang_v1}\includegraphics[width=1.3in]{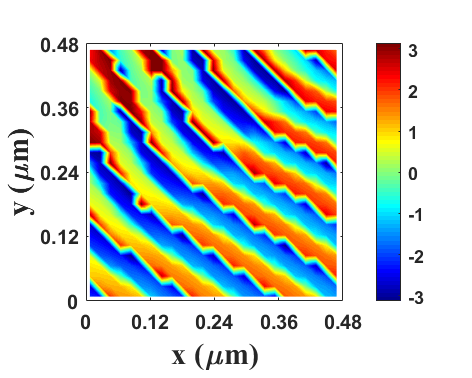}}
	\subfloat{\label{BDF3_alpha_100_ang_v1}\includegraphics[width=1.3in]{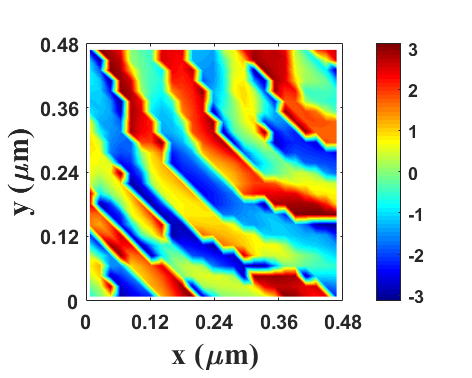}}
	\hspace{0.1in}
	\subfloat{\label{BDF2_alpha_1_ang_v1}\includegraphics[width=1.3in]{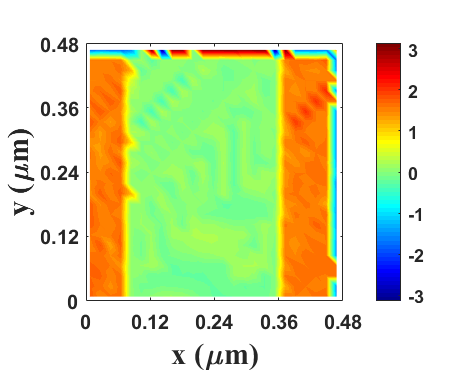}}
	\subfloat{\label{semiBDF2_alpha_5_ang_v1}\includegraphics[width=1.3in]{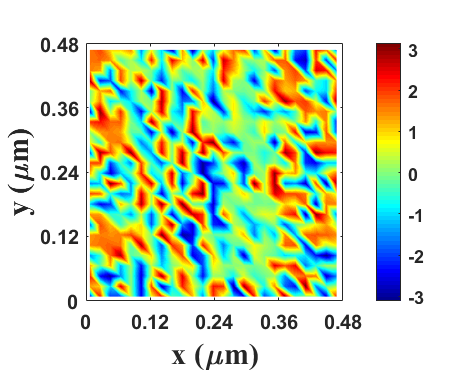}}
	\subfloat{\label{semiBDF2_alpha_10_ang_v1}\includegraphics[width=1.3in]{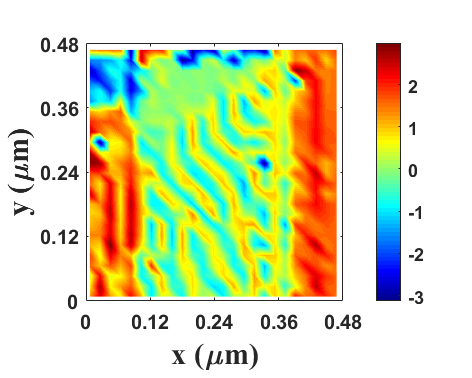}}
	\subfloat{\label{semiBDF2_alpha_40_ang_v1}\includegraphics[width=1.3in]{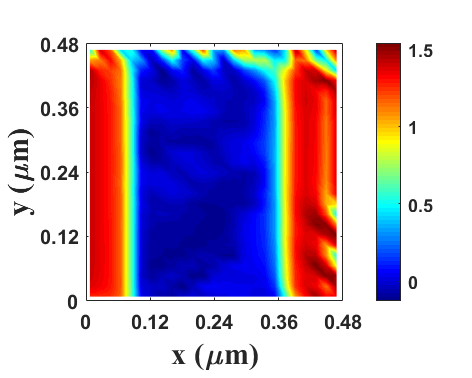}}
	\subfloat{\label{semiBDF2_alpha_100_ang_v1}\includegraphics[width=1.3in]{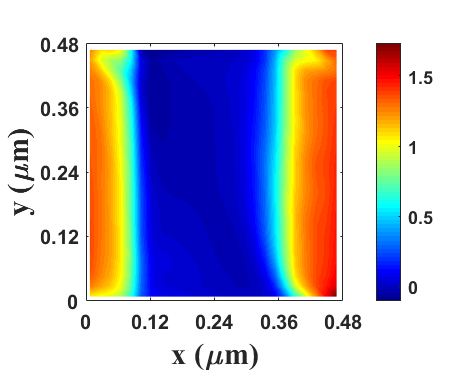}}
	\hspace{0.1in}
	\subfloat{\label{BD1_alpha_1_ang_v1}\includegraphics[width=1.3in]{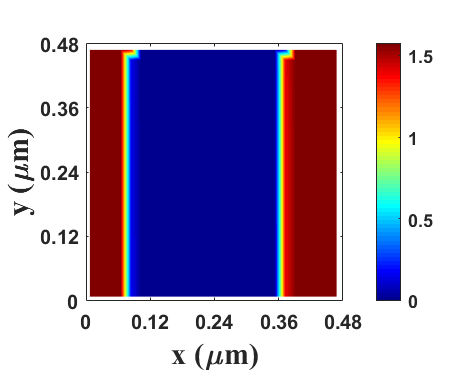}}
	\subfloat{\label{BDF1_alpha_5_ang_v1}\includegraphics[width=1.3in]{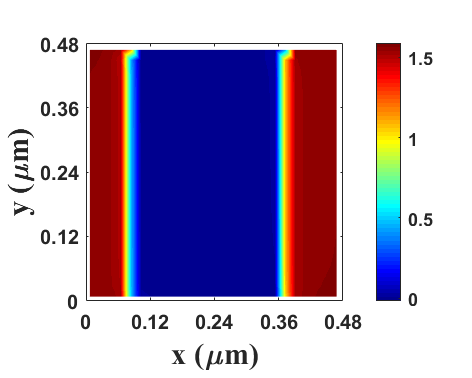}}
	\subfloat{\label{BDF1_alpha_10_ang_v1}\includegraphics[width=1.3in]{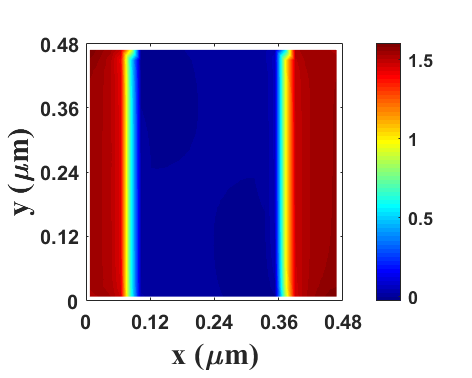}}
	\subfloat{\label{BDF1_alpha_40_ang_v1}\includegraphics[width=1.3in]{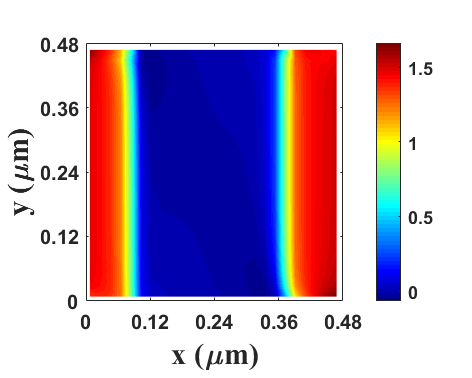}}
	\subfloat{\label{BDF1_alpha_100_ang_v1}\includegraphics[width=1.3in]{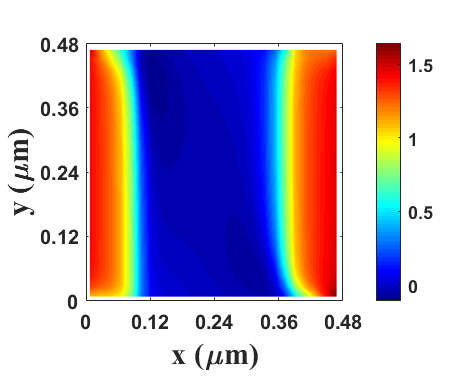}}
	\caption{Stable structures in the absence of magnetic field at $2\,$ns when $\alpha=1,5,10,40,100$. The color denotes the angle between the first two components of the magnetization vector. Top: Proposed method; Middle: BDF2; Bottom: BDF1; From left to right: $\alpha=1,5,10,40,100$, $dt=0.1\;ps$. }\label{BDF3_BDF2_BDF1_alpha_v1}
\end{figure}

To check the numerical stability of these three methods in the practical simulations of micromagnetics with large damping parameters, we consider a thin film of size $480\times480\times20\,\textrm{nm}^3$ with grid points $100\times100\times4$. The temporal step-size is taken as $k=1\,$ps and $k=0.1\,$ps. A uniform state along the $x$ direction is set to be the initial magnetization and the external magnetic field is set to be $0$. Three different damping parameters, $\alpha=1, 5,10,40,100$, are tested with unstable magnetization profiles shown in \cref{BDF3_BDF2_BDF1_alpha} and \cref{BDF3_BDF2_BDF1_alpha_v1}. In particular, the following observations are made. 
\begin{itemize}
	\item The proposed method is unstable and failed for large damping parameters;
	\item As the order of the constructive numerical method increases, the simulation becomes more unstable.;
	\item BDF1 method is hopefully to be stable to capture the physical structure.
\end{itemize}
In fact, a preliminary theoretical analysis reveals that, an optimal rate convergence estimate of the proposed method could be theoretically justified for $\alpha>7$ (BDF2 method with $\alpha>3$).

\begin{figure}[htbp]
	\centering
	\subfloat[Proposed]{\label{energy_BDF3_v1}\includegraphics[width=1.8in]{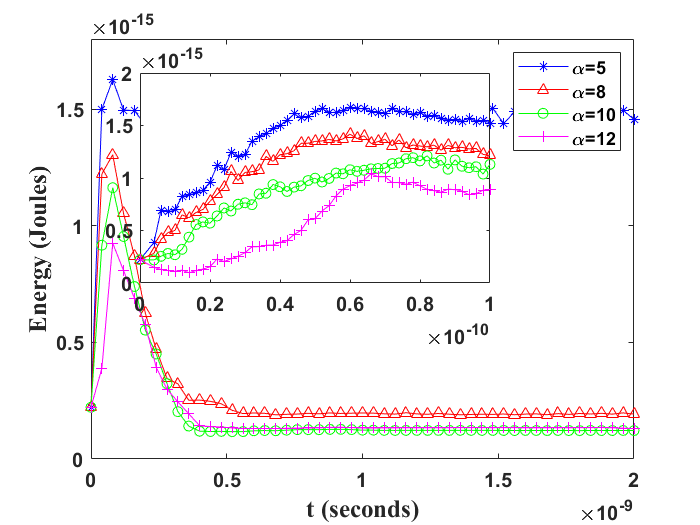}}
	\subfloat[BDF2]{\label{energy_BDF2_v1}\includegraphics[width=1.8in]{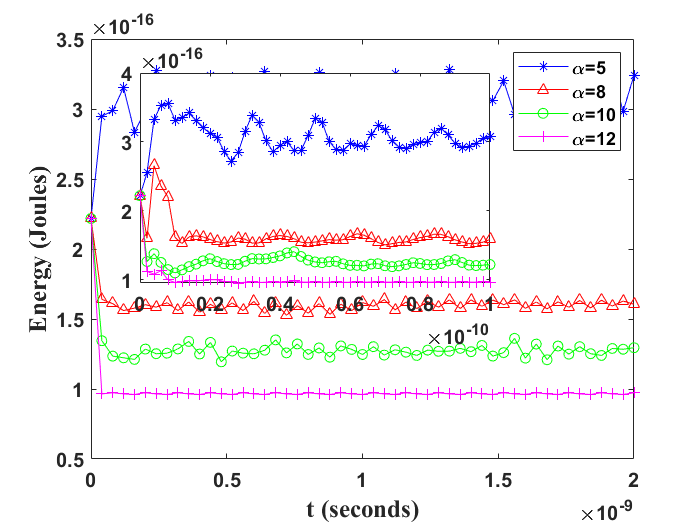}}
	\subfloat[BDF1]{\label{energy_BDF1_v1}\includegraphics[width=1.8in]{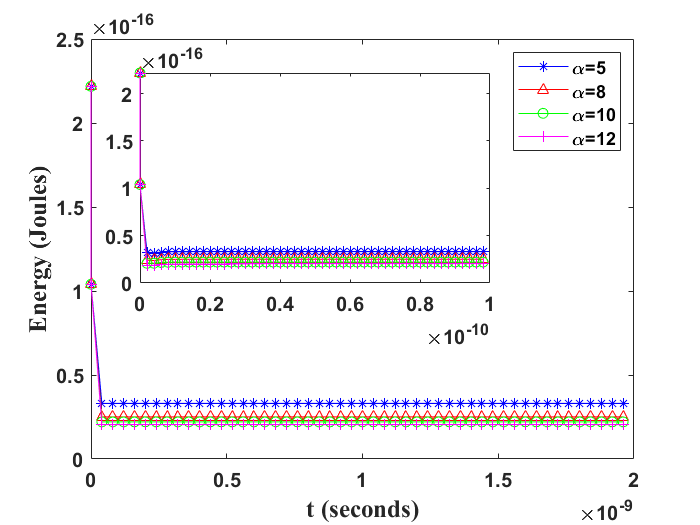}}
	\hspace{0.2in}
	\subfloat[Proposed]{\label{energy_BDF3_v2}\includegraphics[width=1.8in]{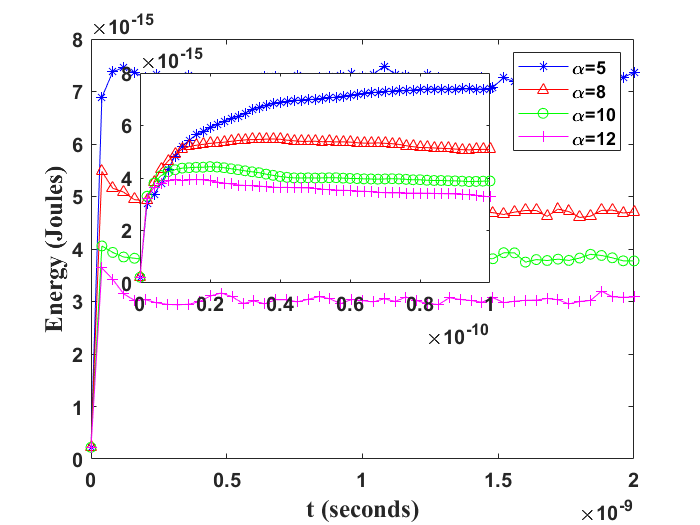}}
	\subfloat[BDF2]{\label{energy_BDF2_v2}\includegraphics[width=1.8in]{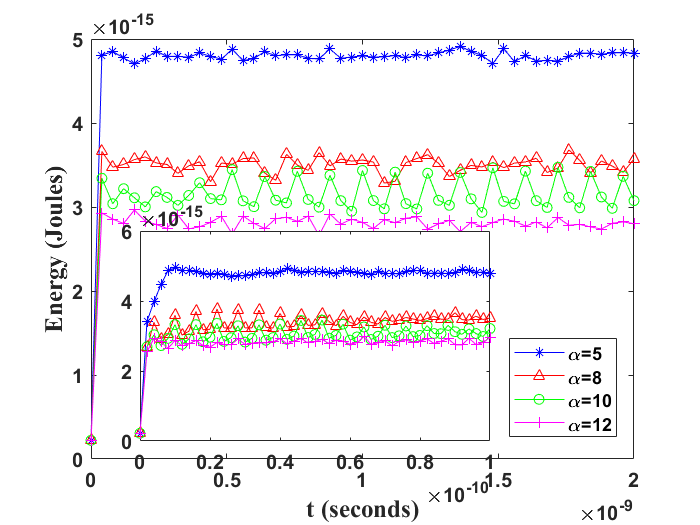}}
		\subfloat[BDF1]{\label{energy_BDF1_v2}\includegraphics[width=1.8in]{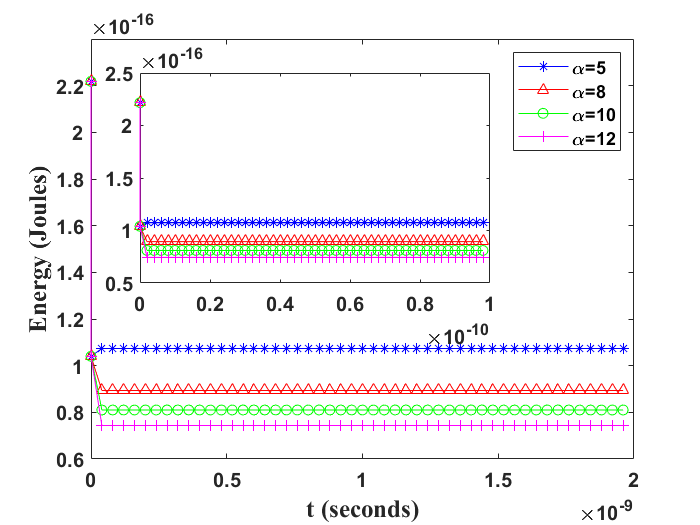}}
	\caption{Energy evolution curves of three numerical methods, with different damping constants, $\alpha=5,8,10,12$, up to $t=2\,$ns  in the absence of external magnetic field. Left: Proposed numerical method; Middle: BDF2; Right: BDF1. One common feature is that the energy dissipation rate is faster for larger $\alpha$. Top row: $dt=1\;ps$; Bottom row: $dt=0.1\;ps$}\label{energy_decay}
\end{figure}

Under the same setup outlined above, we investigate the energy dissipation of the proposed method, the BDF2, and the BDF1. The steady state is attainable at $t=2\,\textrm{ns}$, while the total energy is computed by \eqref{LL-Energy}. The energy evolution curves of different numerical methods with different damping parameters, $\alpha=5,8,10,12$, are displayed in \cref{energy_decay}. One common feature is that the energy dissipation rate turns out to be faster for larger $\alpha$, in all three schemes. Meanwhile, a theoretical derivation also reveals that the energy dissipation rate in the LLG equation \eqref{c1-large} depends on $\alpha$, and a larger $\alpha$ leads to a faster energy dissipation rate.
For the BDF2 and BDF3 numerical methods, the energy shows a rising and unstable trend, while for the BDF1 method, the energy generally decreases. As the order of the method increases, the instability of the total micromagnetic energy becomes more pronounced.

\begin{figure}[htbp]
	\centering
	\subfloat[$\alpha=5$]{\label{alpha_5}\includegraphics[width=2.5in]{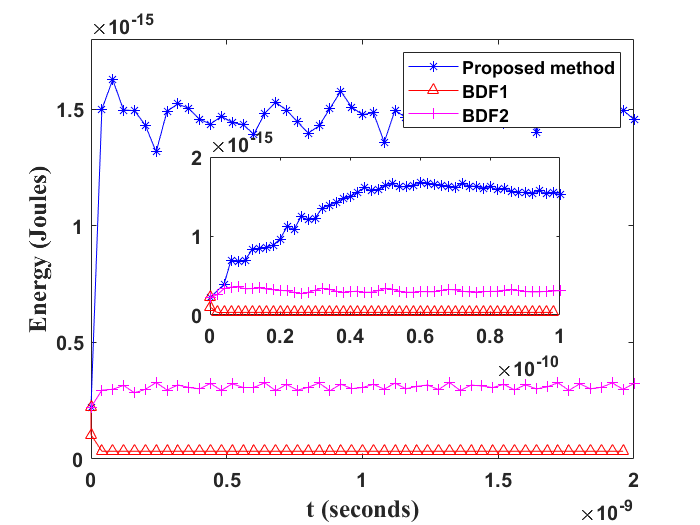}}
	\subfloat[$\alpha=8$]{\label{alpha_8}\includegraphics[width=2.5in]{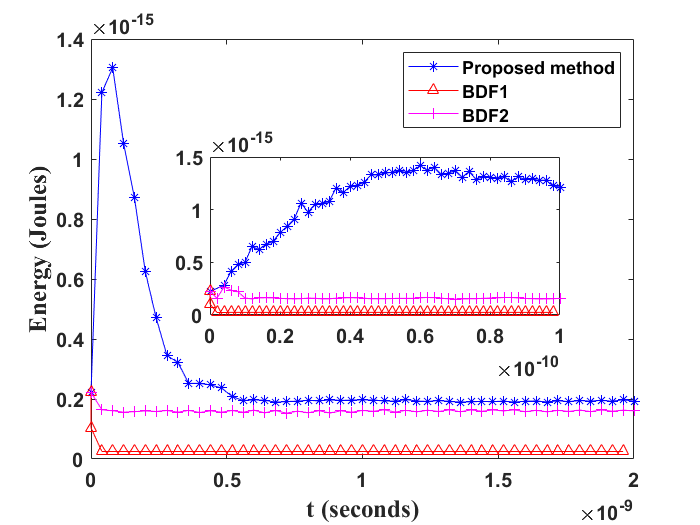}}
	\hspace{0.1in} 
	\subfloat[$\alpha=10$]{\label{alpha_10}\includegraphics[width=2.5in]{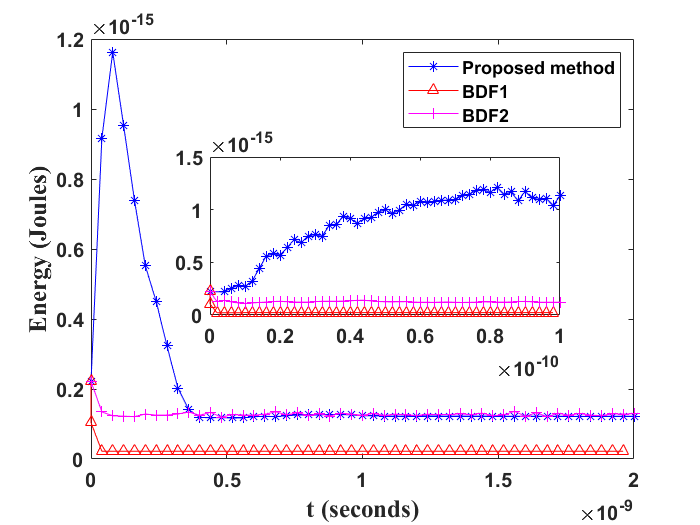}}
	\subfloat[$\alpha=12$]{\label{alpha_12}\includegraphics[width=2.5in]{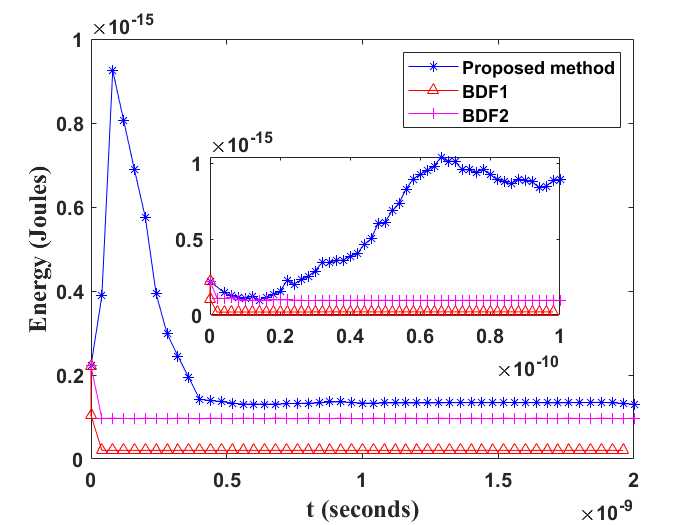}}
	\caption{Energy evolution curves in terms of time, for the numerical results created by three numerical methods up to $t=2\,$ns with $dt=1\;ps$ in the absence of external magnetic field for (a) $\alpha=5$, (b) $\alpha=8$, (c) $\alpha=10$, and (d) $\alpha=12$. The energy dissipation pattern of the proposed method is different with the other two methods, and the BDF1 has a relatively reasonable energy dissipation pattern from the other two methods. As $\alpha$ increases, the stability of the method becomes worse. As $\alpha$ increases, the energy value of the proposed method at the final moment decreases.}\label{energy_decay_alpha}
\end{figure}

\begin{figure}[htbp]
	\centering
	\subfloat[$\alpha=25$]{\label{alpha_5_v1}\includegraphics[width=2.5in]{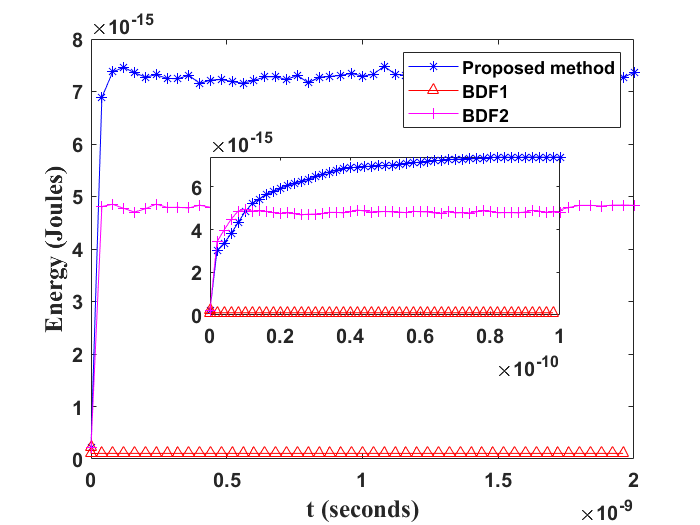}}
	\subfloat[$\alpha=8$]{\label{alpha_8_v1}\includegraphics[width=2.5in]{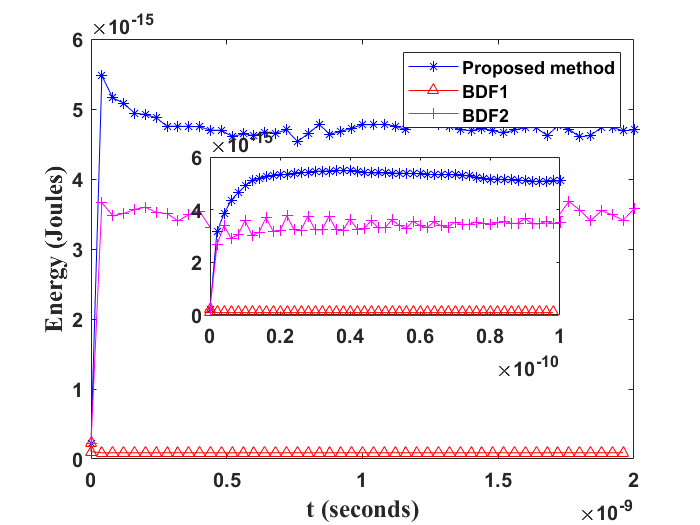}}
	\hspace{0.1in} 
	\subfloat[$\alpha=10$]{\label{alpha_10_v1}\includegraphics[width=2.5in]{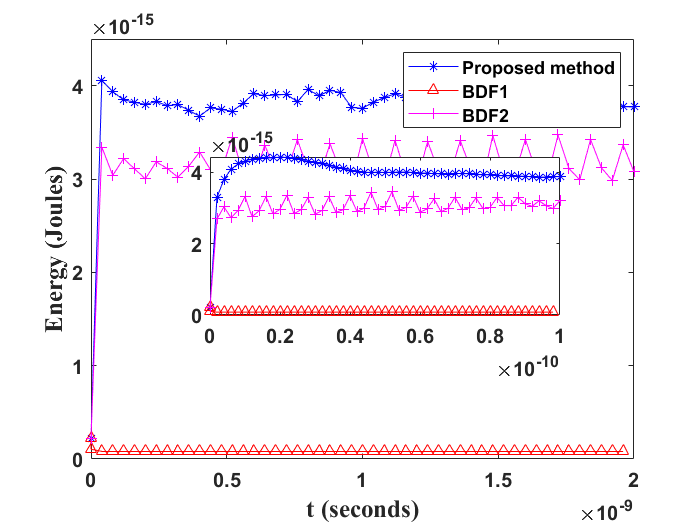}}
	\subfloat[$\alpha=12$]{\label{alpha_12_v1}\includegraphics[width=2.5in]{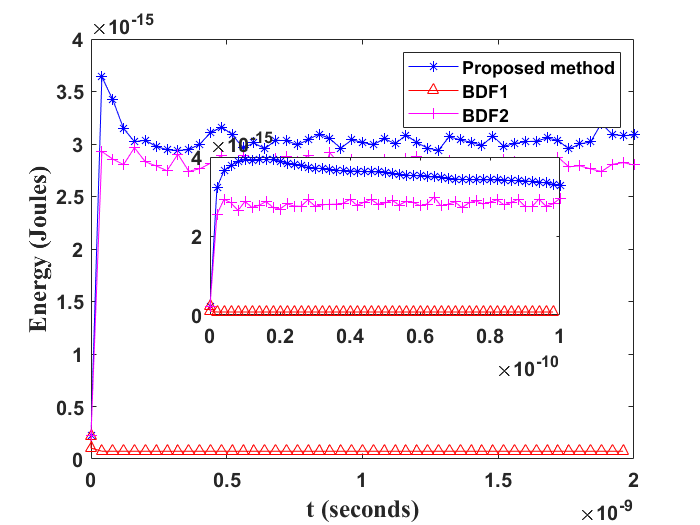}}
	\caption{Energy evolution curves in terms of time, for the numerical results created by three numerical methods up to $t=2\,$ns with $dt=0.1\;ps$ in the absence of external magnetic field for (a) $\alpha=5$, (b) $\alpha=8$, (c) $\alpha=10$, and (d) $\alpha=12$. The energy of BDF2 and the proposed method increases with energy instability, while the energy of BDF1 decreases and remains relatively stable. Even when the time step is sufficiently small at this point, micromagnetic energy instability still occurs. }\label{energy_decay_alpha_v1}
\end{figure}

Meanwhile, we choose the same sequence of values for $\alpha$, and display the energy evolution curves in terms of time up to $T=2\,$ns with $k=1\;$ps in \cref{energy_decay_alpha}. It is found that 
the proposed method is the most unstable, followed by BDF2, while BDF1 is relatively the most stable for $\alpha=5,8,10$ and $12$. As $\alpha$ increases, the energy value of the proposed method at the final moment decreases. Even when the time step is sufficiently small with $k=0.1\;$ps, micromagnetic energy instability still occurs, which is shown in \cref{energy_decay_alpha_v1}

\subsection{Domain wall motion}
\begin{figure}[htbp]
	\centering
	\subfloat[Magnetization for initial state]{\label{NeelWall_initial_mag}\includegraphics[width=2.8in]{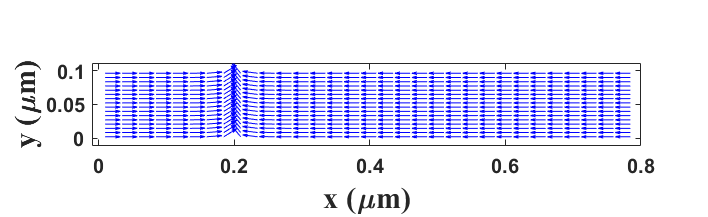}}
		\subfloat[Magnetization for initial state]{\label{NeelWall_initial_mag_v1}\includegraphics[width=2.8in]{BDF3_NeelWall_initial_mag_v1.png}}
	\hspace{0.1in}
	\subfloat[Magnetization with $\alpha=5$  at $2\,$ns]{\label{NeelWall_alpha_2_mag}\includegraphics[width=2.8in]{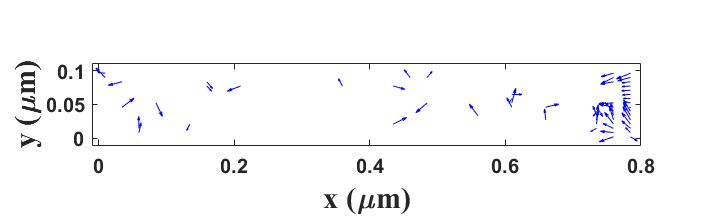}}
	\subfloat[Magnetization with $\alpha=5$  at $2\,$ns]{\label{NeelWall_alpha_2_mag_v1}\includegraphics[width=2.8in]{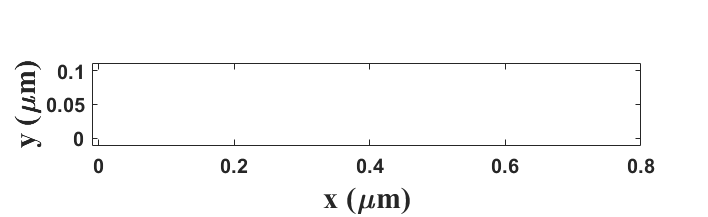}}
	\hspace{0.1in}
	\subfloat[Magnetization with $\alpha=8$ at $2\,$ns]{\label{NeelWall_alpha_8_mag}\includegraphics[width=2.8in]{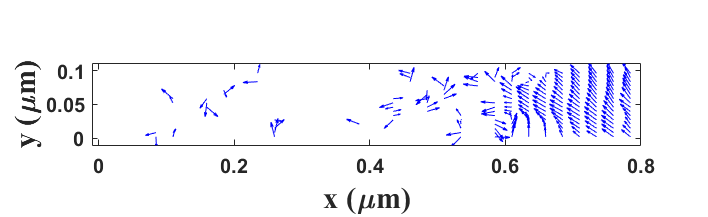}}
	\subfloat[Magnetization with $\alpha=8$ at $2\,$ns]{\label{NeelWall_alpha_8_mag_v1}\includegraphics[width=2.8in]{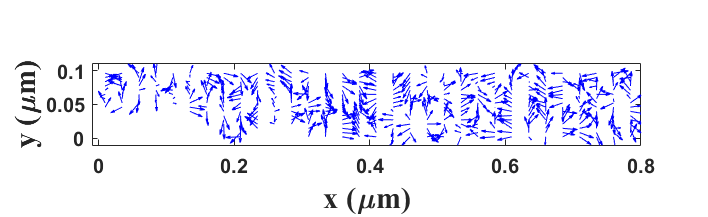}}
	\hspace{0.1in}
	\subfloat[Magnetization with $\alpha=10$ at $2\,$ns]{\label{NeelWall_alpha_10_mag}\includegraphics[width=2.8in]{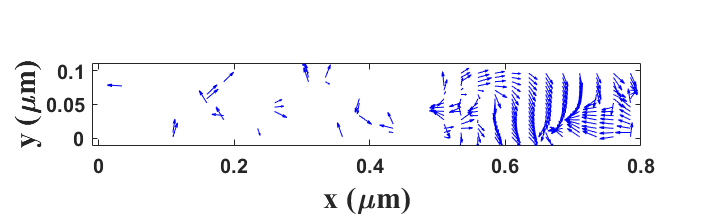}}
	\subfloat[Magnetization with $\alpha=10$ at $2\,$ns]{\label{NeelWall_alpha_10_mag_v1}\includegraphics[width=2.8in]{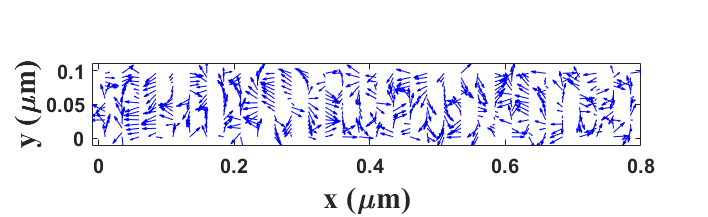}}
	\caption{Magnetization profiles of Ne\'{e}l wall motion in the presence of a magnetic field $\h_e=5\,$mT, with $\alpha = 5,8,10$ at $2\,$ns for the proposed numerical method. The in-plane arrow denotes the first two components of the magnetization vector. Left panel: $k=1\;ps$; Right panel: $k=0.1\;ps$. }\label{NeelWall_alpha_2ns}
\end{figure}

\begin{figure}[htbp]
	\centering
	\subfloat[Magnetization for initial state]{\label{NeelWall_initial_mag_v2}\includegraphics[width=2.8in]{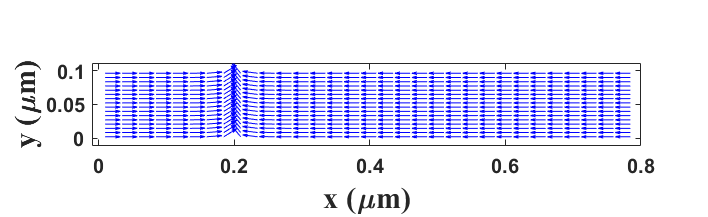}}
	\subfloat[Magnetization for initial state]{\label{NeelWall_initial_mag_v3}\includegraphics[width=2.8in]{BDF2_NeelWall_initial_mag_v2.png}}
	\hspace{0.1in}
	\subfloat[Magnetization with $\alpha=5$  at $2\,$ns]{\label{NeelWall_alpha_2_mag_v2}\includegraphics[width=2.8in]{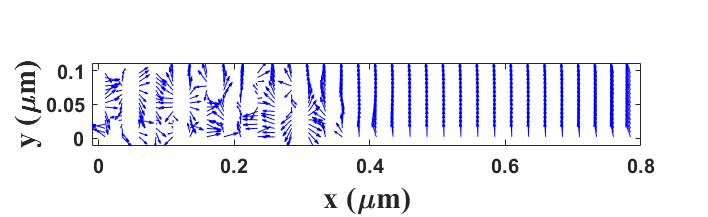}}
	\subfloat[Magnetization with $\alpha=5$  at $2\,$ns]{\label{NeelWall_alpha_2_mag_v3}\includegraphics[width=2.8in]{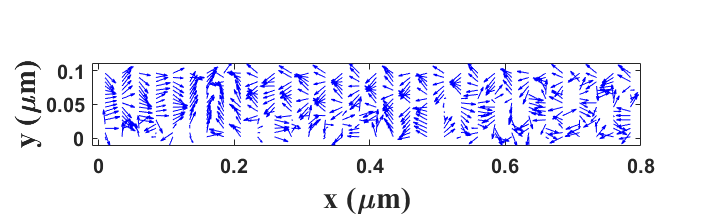}}
	\hspace{0.1in}
	\subfloat[Magnetization with $\alpha=8$ at $2\,$ns]{\label{NeelWall_alpha_8_mag_v2}\includegraphics[width=2.8in]{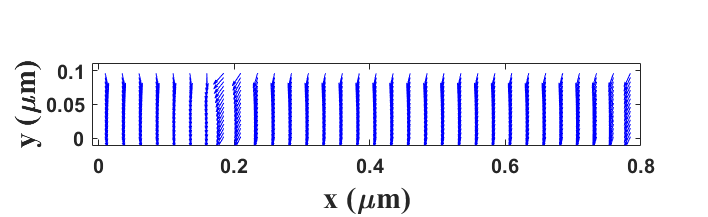}}
	\subfloat[Magnetization with $\alpha=8$ at $2\,$ns]{\label{NeelWall_alpha_8_mag_v3}\includegraphics[width=2.8in]{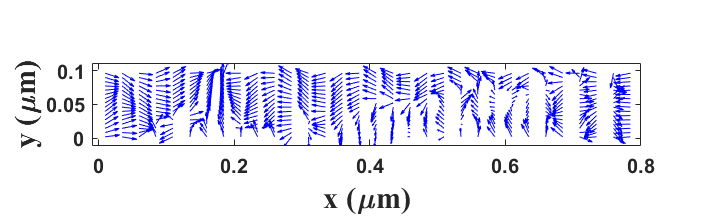}}
	\hspace{0.1in}
	\subfloat[Magnetization with $\alpha=10$ at $2\,$ns]{\label{NeelWall_alpha_10_mag_v2}\includegraphics[width=2.8in]{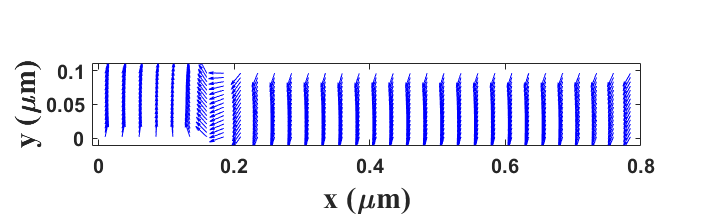}}
	\subfloat[Magnetization with $\alpha=10$ at $2\,$ns]{\label{NeelWall_alpha_10_mag_v3}\includegraphics[width=2.8in]{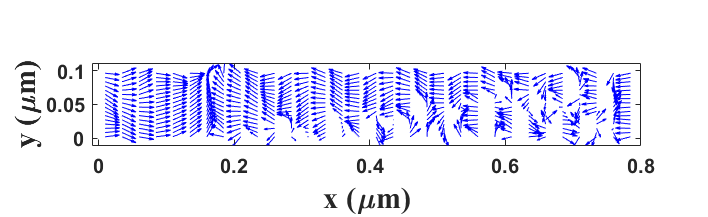}}
	\caption{Magnetization profiles of Ne\'{e}l wall motion in the presence of a magnetic field $\h_e=5\,$mT, with $\alpha = 5,8,10$ at $2\,$ns for the BDF2 method. The in-plane arrow denotes the first two components of the magnetization vector. Left panel: $k=1\;ps$; Right panel: $k=0.1\;ps$. }\label{NeelWall_alpha_2ns_v1}
\end{figure}

\begin{figure}[htbp]
	\centering
	\subfloat[Magnetization for initial state]{\label{NeelWall_initial_mag_v4}\includegraphics[width=2.8in]{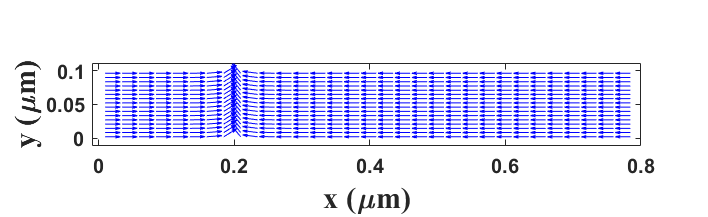}}
	\subfloat[Magnetization for initial state]{\label{NeelWall_initial_mag_v5}\includegraphics[width=2.8in]{BDF1_NeelWall_initial_mag_v2.png}}
	\hspace{0.1in}
	\subfloat[Magnetization with $\alpha=5$  at $2\,$ns]{\label{NeelWall_alpha_2_mag_v4}\includegraphics[width=2.8in]{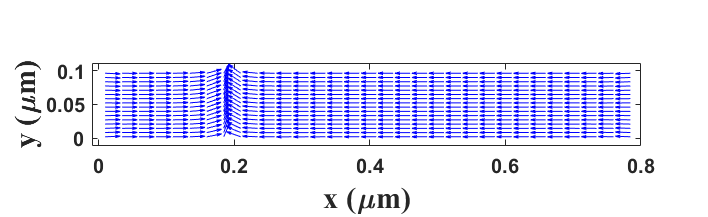}}
	\subfloat[Magnetization with $\alpha=5$  at $2\,$ns]{\label{NeelWall_alpha_2_mag_v5}\includegraphics[width=2.8in]{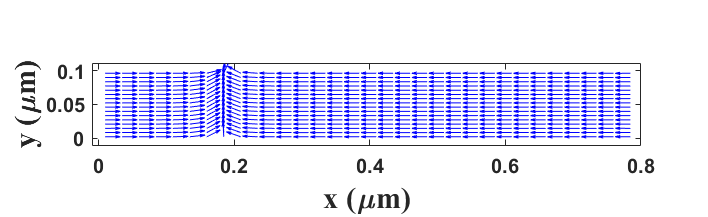}}
	\hspace{0.1in}
	\subfloat[Magnetization with $\alpha=8$ at $2\,$ns]{\label{NeelWall_alpha_8_mag_v4}\includegraphics[width=2.8in]{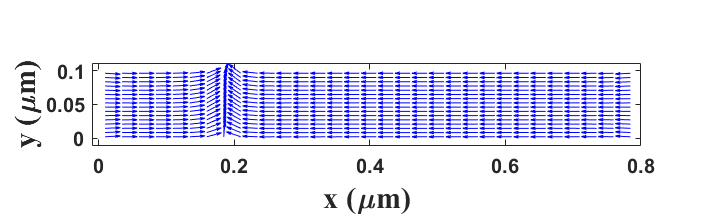}}
	\subfloat[Magnetization with $\alpha=8$ at $2\,$ns]{\label{NeelWall_alpha_8_mag_v5}\includegraphics[width=2.8in]{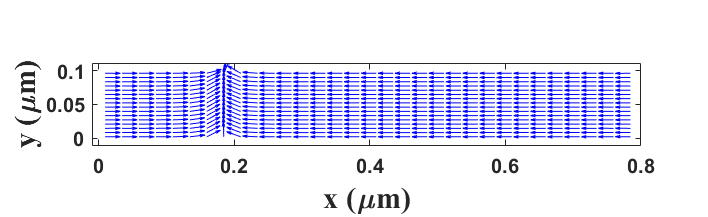}}
	\hspace{0.1in}
	\subfloat[Magnetization with $\alpha=10$ at $2\,$ns]{\label{NeelWall_alpha_10_mag_v4}\includegraphics[width=2.8in]{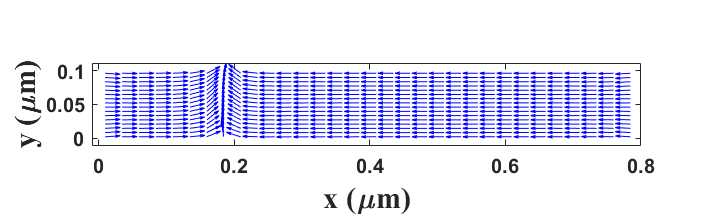}}
	\subfloat[Magnetization with $\alpha=10$ at $2\,$ns]{\label{NeelWall_alpha_10_mag_v5}\includegraphics[width=2.8in]{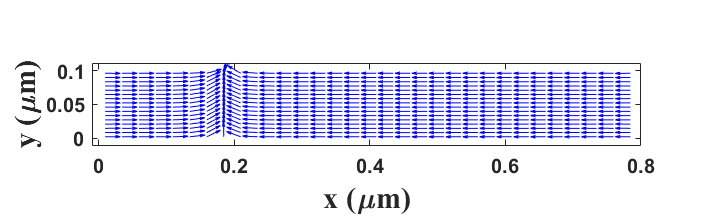}}
	\caption{Magnetization profiles of Ne\'{e}l wall motion in the presence of a magnetic field $\h_e=5\,$mT, with $\alpha = 5,8,10$ at $2\,$ns for the BDF1 method. The in-plane arrow denotes the first two components of the magnetization vector. Left panel: $k=1\;ps$; Right panel: $k=0.1\;ps$. }\label{NeelWall_alpha_2ns_v2}
\end{figure}

A Ne\'el wall is initialized in a nanostrip of size $800\times100\times4\,\textrm{nm}^3$ with grid points $128\times64\times4$. An external magnetic field of $\h_e=5\,$mT is then applied along the positive $x$ direction and the domain wall dynamics is simulated up to $2\,$ns with $\alpha=2,5,8$. The corresponding magnetization profiles are visualized in \cref{NeelWall_alpha_2ns}, \cref{NeelWall_alpha_2ns_v1} and \cref{NeelWall_alpha_2ns_v2}. It is observed that the domain wall of the BDF3 method fails, and invalid magnetization is obtained, with the method still exhibiting strong instability. The domain wall of the BDF2 method also fails, and the method shows a certain degree of instability; however, this instability seems to improve as $\alpha$ increases. Even when a small time step is adopted, this instability still persists. The BDF1 method appears to be the most stable; a certain tendency of domain wall movement is observed, but this movement state does not change significantly as $\alpha$ increases.

\section{Conclusions}
\label{sec:conclusions}

This paper develops several accurate numerical methods for solving the Landau-Lifshitz-Gilbert equation under large damping parameters, focusing on their stability in micromagnetics simulations. The schemes integrate backward-differentiation formula for temporal discretization, implicit handling of the constant-coefficient diffusion term, and fully explicit extrapolation for nonlinear terms (including the gyromagnetic term and the nonlinear segment of the harmonic mapping flow). Leveraging large damping, the methods are shown to be unconditionally stable. Numerical experiments in 1D and 3D validate their accuracy and efficiency. Furthermore, micromagnetics simulations comparing the proposed method with two other approaches reveal that pre-projected solutions introduce instability in practical applications, a finding corroborated by energy dissipation and domain wall dynamics tests.

\section*{Acknowledgments}
This work is supported in part by the grants NSF DMS-2012669 (C.~Wang), and Jiangsu
Science and Technology Programme-Fundamental Research Plan Fund, Research and Development
Fund of XJTLU (RDF-24-01-015) (C. Xie)

\bibliographystyle{amsplain}
\bibliography{references}

\end{document}